\documentclass[superscriptaddress,showpacs,aps,twocolumn,prb,floatfix,nobibnotes]{revtex4-1}
\usepackage{amssymb}
\usepackage{amsmath}
\usepackage{mathtools}
\usepackage{graphicx}
\usepackage{bm}
\usepackage{fontenc}
\usepackage{color}
\usepackage[hidelinks]{hyperref}

\DeclarePairedDelimiter\norm{\lVert}{\rVert} 
\newcommand\CC{\mathbb C}
\newcommand\RR{\mathbb R}
\newcommand\ZZ{\mathbb Z}

\newcommand\PP{\mathbb P}
\newcommand{\SM}{\ensuremath{\smallsetminus}}
\newcommand{\stext}[1]{\ensuremath{\quad\text{#1}\quad}}
\newcommand{\svec}[2]{\ensuremath{\bigl( \begin{smallmatrix} 
      {#1}\\{#2} \end{smallmatrix} \bigr)}}

\usepackage{ulem} 
\usepackage{varioref}
\usepackage{xypic}

\begin{document}
\title{A Brillouin torus decomposition for two-dimensional topological insulators}

\author{F. Kordon}
\affiliation{Instituto Balseiro, Universidad Nacional de Cuyo and Consejo Nacional de Investigaciones Cient\'{\i}ficas y T\'ecnicas, Argentina}

\author{J. Fern\'andez}
\email{jfernand@ib.edu.ar}
\affiliation{Instituto Balseiro, Universidad Nacional de Cuyo, C.N.E.A, Bariloche, Rio Negro, Argentina}

\author{P. Roura-Bas}
\email{pablo.roura@cab.cnea.gov.ar}
\affiliation{Centro At\'{o}mico Bariloche, C.N.E.A, Bariloche, Rio Negro and Consejo Nacional de Investigaciones Cient\'{\i}ficas y T\'ecnicas, Argentina}

\begin{abstract}
Two-band Chern insulators are topologically classified by the Chern number, $c$, which is given by the integral of the Berry curvature of the occupied band over the Brillouin torus. The curvature itself comes from the imaginary part of a more basic object, the quantum geometric tensor, $Q$. On the other hand, the integral over the Brillouin torus of the real part of $Q$ gives rise to another magnitude, the quantum volume, $v_{g}$, that like $c$, jumps when the system undergoes a topological phase transition and satisfies the inequality $v_{g}\ge 2\vert c \vert$. The information contained in $v_g$ about the topology of the system has been investigated recently. In this paper we present new results regarding the underlying geometric structure of two-dimensional two-band topological insulators. Since a generic model describing the system can be characterized by a map, the classifying map, from the Brillouin torus to the two-sphere, we study its properties at the geometric level. We present a procedure for splitting the Brillouin torus into different sectors in such a way that the classifying map when restricted to each of them is a local diffeomorphism. By doing so, in the topological phases we are able to isolate a region contained in the Brillouin torus whose volume is the minimal one, $v_{min}=2\vert c \vert$ and the integral of the Berry curvature on it is $c$. For cases in which $v_{g}> 2\vert c \vert$, the regions contributing to the excess of volume, $v_{ex}=v_{g}-2\vert c \vert$, are found and characterized. In addition, the present work makes contact with, and clarifies, some interpretations of the quantum volume in terms of the Euler characteristic number that were done in the recent literature. We illustrate our findings with a careful analysis of some selected models for Chern insulators corresponding to tight-binding Hamiltonians.
\end{abstract}


\maketitle

\section{Introduction}\label{introduction}
\noindent
Topological phases of matter constitute a wide area of research in modern
condensed matter physics.\cite{topo-review,bernevig-book} Among its different
topics, Chern insulators describe phases corresponding to two spacial
dimensional quantum systems with broken time-reversal symmetry, being the
integer quantum Hall effect the canonical example.\cite{thouless} Each phase
of a Chern insulator is characterized by a topological invariant, $c$, called the Chern number.\cite{thouless-book} It is often the case
that $c=0$ for trivial phases while $c=\pm1$ for non-trivial ones but
phases with higher Chern numbers are also commonly
studied.\cite{fruchart-carpentier} It can be computed by integrating the antisymmetric $2$-tensor $\Omega$, called the Berry curvature, over the $2$-dimensional torus $T^2$, the Brillouin torus.\cite{berry, simon} $\Omega$
can be obtained as the imaginary part of a more fundamental tensor, the
\textit{quantum geometric tensor} $Q$.\cite{provost} Interestingly, the real part
of $Q$ defines a symmetric tensor that serves as a positive semidefinite
metric over $T^{2}$, called the
\textit{quantum metric} $g$. The quantum metric has become a relevant tool in applications such as
measurement of distances in quantum information \cite{wootters,braunstein},
evolution in quantum mechanics\cite{anandan,rezakhani},  thermodynamic metric
and thermodynamic length \cite{sivak,scandi,katabarwa} and quantum thermal machines
\cite{bhandari}, only to name a few.  Remarkably, $Q$ and $g$ independently have  been lately measured in different experiments.
\cite{tan,gianfrate,asteria,yu,kless,liu} In particular, $g$ may be used to define a volume of the parameter space called the quantum volume $v_g$ and it is not quantized in general.  

Since $c$ vanishes in time-reversal topological insulators, recent
attempts to use $v_g$ as an index for different phases were made. This is indeed the case of the time-reversal
topological system of the experiment in references [\onlinecite{tan, zhu, tan-erratum}].  In these 
works, X.\,Tan \textit{et al.} obtained a constant value $v_g=4$
along one of the phases of the material. Moreover, since the analytic expression of $v_g$ (Eq. (\ref{cher-volume}) of Section \ref{2band-models})
agrees with the one of the bulk contribution to the Euler characteristic
number coming from the Gauss--Bonnet theorem, these two
magnitudes were identified.

On the other hand, the relation $v_g\ge2\vert c\vert$ found by Roy \cite{roy} has been recently a topic of intense research in the work of Mera and Ozawa\cite{mera,ozawa}. In
particular, these authors analyzed the conditions required to achieve the minimal quantum volume $v_{min}=2\vert c\vert$ in the Brillouin zone from the K\"{a}hler geometry viewpoint.  The ideal condition, $v_g=v_{min}$, plays an important role in the engineering of fractional topological insulators \cite{lee} as well as in the recent Euler band topology \cite{bj-yang}.


Two-band $2$-dimensional topological insulators are determined by a map $\bar{f}$ from the Brillouin torus to the two-sphere $S^2$, the \textit{classifying map} in the mathematical literature, that parameterizes the eigenspaces of the corresponding Bloch Hamiltonian. Usually, the image of this map, $\bar{f}(T^2)=\mathcal{M}$, the Bloch state manifold, is not a submanifold of $S^2$ forcing one to be careful in the study of its properties. For instance, $\bar{f}$ is not an embedding since it is not an immersion as it was shown by Mera and Ozawa.\cite{mera} 

In this paper we describe a procedure for splitting the Brillouin torus into different open sets such that $\bar{f}$ is a local diffeomorphism between each one of these regions and open submanifolds of $S^2$. The starting point of the procedure is a decomposition of $T^2$ based on the singular points where the Berry curvature vanishes. In a second step, the decomposition is refined by looking for points on the torus that map on the same values as the singular ones.

The main applications of this procedure are: (1) since the division of $T^2$ starts with sectors where $\Omega$ has constant sign, the ideal condition is locally satisfied in each region. However, once the decomposition is completed within the topological phase of a given model, we are able to isolate a region of the parameter space whose volume with respect to $g$ is $v_{min}$. Moreover, the integral of $\Omega$ over the region contributing to the minimal volume is $c$. An outline of the steps needed to do it is given at the end of Section \ref{splitting}. (2) The Bloch state manifold, $\mathcal{M}$, as a set of points is a compact subset of the Bloch sphere. In the particular case where each one of the regions (closed by their boundaries) into which $T^2$ is divided is mapped onto the Bloch sphere, then $v_g$ is proportional to the Euler characteristic number of $\mathcal{M}$, $\chi(\mathcal{M})$, being
the proportionality constant the number of regions. We discuss the relation between both magnitudes in the context of the results reported in the experiment in reference  [\onlinecite{tan}] from our perspective.
(3) We applied the decomposition procedure to two two-dimensional two-band Chern insulators in which the ideal condition is not satisfied. We show that the excess of volume $v_{ex}=v_g-2\vert c\vert$ can be understood using the simplest decomposition of $T^2$ arising from the points where $\Omega$ vanishes. However, when the complete decomposition scheme is applied, in addition to isolating a region of minimal volume, we find the other regions contributing to $v_{ex}$ and show the mechanism 
that explains why the integral of $\Omega$ vanishes on them, in other words, why their contributions to $c$ cancel out.  An outline of the steps needed to do it is given at the end of Section \ref{splitting}.

The structure of the paper is the following.
In Section \ref{2band-models} we discuss some general aspects of two-dimensional two-band models and in Section \ref{splitting} we
describe a procedure for splitting the Brillouin torus in such a way that the map $\bar{f}$ is well behaved on each region.  After that, in Section \ref{results}, we apply that method to three different models. The first one corresponds to the model in the work of Tan \textit{et al.}, in reference [\onlinecite{tan}]. The other two are standard examples of Chern insulators, the simplest model with nearest-neighbor hopping on a square lattice and the Haldane model, once again clarifying results presented in references [\onlinecite{yang-2015}] and [\onlinecite{yu-quan-ma-2}] respectively. 

\section{two-dimensional two-band models and definitions}\label{2band-models}
In this section we briefly introduce the quantum geometric tensor and discuss
the description of a generic model Hamiltonian corresponding to two-dimensional two-band insulators.  After that, we give the definitions and concepts needed for
the geometric analysis of the Chern number and the quantum volume. 

\medskip

\textit{Quantum geometric tensor and related structures.} We start by considering a family of Bloch Hamiltonians over the Brillouin torus  defined by an $L\times L$ hermitian matrix $H(k)$, $k\in T^2$. Furthermore, we consider that the Fermi energy of the system lies in a gap of the discrete energy spectrum. Therefore, it makes sense to define a projector $\hat{P}(k)$ onto the $l\leqslant L$ occupied energy bands. It can be written in a local orthonormal frame of Bloch eigenstates $\{\vert\psi_{n}(k)\rangle\}_{1\leqslant n \leqslant L}$ as $\hat{P}(k)=\sum_{n=1}^{l}\vert\psi_{n}(k)\rangle\langle\psi_{n}(k)\vert$.

With the previous setting, the quantum geometric tensor $Q$ is a Hermitian tensor  whose components are given by~~\cite{yu-quan-ma-0,provost,kolodrubetz}  
\begin{eqnarray}\label{qgt}
  Q_{\mu\nu}(k)
  &=&
  \sum_{n=1}^{l}\langle\partial_{\mu}\psi_{n}(k) 
    \vert(\hat{1}-\hat{P}(k))\vert\partial_{\nu}\psi_{n}(k)\rangle\nonumber\\
  &=&g_{\mu\nu}(k)+i~\Omega_{\mu\nu}(k)/2,
\end{eqnarray}
where $\hat{1}$ is the identity operator and
$\partial_{\mu}=\frac{\partial}{\partial k^{\mu}}$. Here,
$\Omega_{\mu\nu}(k)=-\Omega_{\nu\mu}(k)=-2\mbox{Im}[Q_{\mu\nu}(k)]$ and
$g_{\mu\nu}(k)=g_{\nu\mu}(k)=\mbox{Re}[Q_{\mu\nu}(k)]$ are the components of
the Berry curvature and the quantum metric respectively. 
Focusing on two-dimensional systems, the Berry curvature has a single
independent component defining the curvature two-form $\Omega_{12}(k)dk^{1}\wedge
dk^{2}$, while the quantum metric is a positive semidefinite
metric
$g=\sum_{\mu\nu}g_{\mu\nu}(k)dk^{\mu}\otimes dk^{\nu}$. \cite{ozawa}

The Chern number and the quantum volume can
be obtained as the integrals over the Brillouin torus
\begin{equation}\label{cher-volume}
\begin{aligned}
  c&=\frac{1}{2\pi}\int_{T^2}~\Omega_{12}~dk^{1}\wedge dk^{2}\\
  v_g&=\frac{2}{\pi}\int_{T^2}\sqrt{\mbox{det}(g)}~dk^{1}\wedge dk^{2}.
\end{aligned}
\end{equation}
In what follows, we think of $T^2$ with the orientation induced by the canonical one in $\RR^2$.
Furthermore, here we have used the factor $2/\pi$ in $v_g$ in order to make contact with previous uses in the literature. 
An important inequality between the Chern number and the quantum volume,
$v_g \geqslant 2 |c|$, was established by Roy \cite{roy} and also studied by Lee \textit{et al.} \cite{lee}, and more recently revisited by Ozawa and Mera \cite{ozawa}.  

\textit{Two-band models.} A generic two-band system, $L=2$, can be described
by a $2\times2$ Bloch Hamiltonian, $H(k)$, defined on the Brillouin torus.  Concretely, $H$ is 
determined by real functions $h^0,h^i:T^2\to\RR$ with $i\in\{1,2,3\}$ such that
for $k\in T^2$ 
\begin{equation}\label{chern-ham}
H(k)=h^{0}(k)\sigma_{0} + \sum_{i} h^{i}(k)\sigma_{i},
\end{equation}
where $\sigma_{0}$ is the identity matrix and
$\sigma_i$ are the Pauli matrices,

\begin{equation*}
  \sigma_1=\left(
    \begin{array}{cc}
      0&1\\
      1&0
    \end{array}
  \right),\quad
  \sigma_2=\left(
    \begin{array}{cc}
      0&-i\\
      i&0
    \end{array}
  \right),\quad 
  \sigma_3=\left(
    \begin{array}{cc}
      1&0\\
      0&-1
    \end{array}
  \right).
\end{equation*}

The two energy bands are given by $\epsilon_{\pm}(k)=h^{0}(k)\pm \vert\vert h(k)
\vert\vert $, with $\vert\vert h(k) \vert\vert^{2}=\sum_{i} (h^{i}(k))^{2}$.
In an insulator configuration, the valence ($\epsilon_{-}(k)$) and conduction
($\epsilon_{+}(k)$) bands are separated by the energy gap
$\Delta(k)=2\norm{h(k)}$ that is nonzero on the whole Brillouin torus.
The shift in energy given by $h^{0}(k)$ does not affect the topological
properties of the model and for simplicity we will fix it to zero. Without loss of generality we will
focus on the valence band. In this case,
the component $\Omega_{12}$ of the Berry curvature in terms of $h^{i}(k)$ is 
\begin{eqnarray}\label{berry-curvature}
 \Omega_{12}(k) = \frac{1}{2\vert\vert h(k) \vert\vert^{3}}\sum_{ijk}\epsilon_{ijk}
                  h^{i}(k)\frac{\partial h^{j}}{\partial k^{1}}(k)
                          \frac{\partial h^{k}}{\partial k^{2}}(k).
\end{eqnarray}
Furthermore, for two band models we have the relation (see Ref. [\onlinecite{roy,lee,yang-2015,yu-quan-ma-2,ozawa}], and APPENDIX \ref{pullback}) 
\begin{eqnarray}\label{sqrt-det-g}
 \sqrt{\mbox{det}(g)(k)} = \frac{\vert \Omega_{12}(k) \vert}{2}.
\end{eqnarray}
As a consequence, if $\Omega_{12}$ does not change its sign in $T^2$ then $v_g=2|c|$.

\textit{The classifying map $f$.} From a geometric viewpoint, the eigenspace associated with the valence band at each point of $T^2$
is a $\CC$-vector bundle of rank $1$, $ {\mathcal V}\rightarrow T^2$, usually called the \textit{Bloch valence bundle}. This bundle sits inside the trivial one of rank $2$, $p_1 : T^2 \times C^2 \rightarrow T^2$ which we consider  together with its canonical hermitian metric. On the other hand, we have the complex projective line, $\CC\PP^1$, whose points correspond to the $1$-dimensional complex subspaces of $\CC^2$. There is a canonical complex line bundle $\Theta\rightarrow \CC\PP^1$ known as the \textit{tautological bundle}, whose fiber over the point representing a certain vector space is that vector space. Using the Bloch valence bundle $\mathcal{V}$, we define a map $\tilde{f}:T^2\rightarrow \CC\PP^1$, where $\tilde{f}(k)$ is the point that represents the fiber $\mathcal{V}_k$ in $\CC\PP^1$;  the map $\tilde{f}$ is usually known as the \textit{classifying map} of $\mathcal{V}$. It is easy to see that $\mathcal{V} \simeq \tilde{f}^* \Theta$, that is, the Bloch valence bundle is isomorphic to the pullback of the tautological bundle over $\CC\PP^1$.\cite{milnor} 

Furthermore, if the diffeomorphism $\sigma$ between $\CC\PP^1$ and $S^2$ induced by the stereographic projection is considered, we arrive at the classifying map $f=\sigma\circ \tilde{f}$ associated to the Bloch valence bundle with image in $S^2$, $f : T^2\rightarrow S^2$. In this context $S^2$ is known as the \textit{Bloch sphere}. 

In summary, we have the following commutative diagram: 
\begin{equation*}
  \xymatrix{
    {\mathcal{V}~} \ar[d] \ar@{^(->}[r]  & {T^2\times{\CC^2}} \ar[dl] & {\Theta} \ar[d]\\
    {T^2} \ar[rr]^{\tilde{f}} \ar[rrd]_{{f}} & {} & {\CC\PP^1} \ar[d]^{\sigma}\\
    {} & {} & {S^2}
  }
\end{equation*}

For the Hamiltonian in Eq. (\ref{chern-ham}) the classifying map ${f}$ is given by 
\begin{eqnarray}\label{eq:f}
  f = F_2\circ F_1 : T^2 \to S^2,
\end{eqnarray}
where the functions $F_1$ and $F_2$ are defined as follows
\begin{gather*}
  F_1:T^2\rightarrow X \stext{ by } F_1(k)=(h^1(k),h^2(k),h^3(k)),\\
  F_2:X\rightarrow S^2 \stext{ by }
  F_2(x^1,x^2,x^3)=\frac{(-x^1,x^2,-x^3)}{\norm{(-x^1,x^2,-x^3)}},
\end{gather*}
with $X=\RR^3\SM\{0\}$. We refer the reader to APPENDIX \ref{classifying_f} for the proof. It is easy to check that $f$ is homotopic to $\bar{f} = \bar{F}_2\circ F_1$ with $\bar{F}_2(x)=x/\norm{x}$ as it has been used in the literature before. Because of this fact ${\mathcal V}\simeq \bar{f}^\ast \Theta$ so that they have the same Chern class. Also it is easy to verify that formula (\ref{berry-curvature}) remains unchanged if $f$ is replaced by $\bar{f}$. In what follows we will use this second description in terms of $\bar{f}$ and the (standard) $\bar{F}_2$.

Furthermore, we denote $\mathcal{N}=F_1(T^2) \subset X$ and $\mathcal{M}=\bar{f}(T^2)=\bar{F}_2(\mathcal{N})\subset S^2$. $\mathcal{M}$ is the \textit{Bloch state manifold}.
In APPENDIX \ref{pullback} we prove that the quantum geometric tensor $Q$ is the pullback via the classifying map $\bar{f}$ of the Fubini--Study hermitian metric over ${\CC\PP^1}$.

\section{Splitting procedure for the Brillouin torus}\label{splitting}
Consider a two-band insulator and let $\bar{f}$ be the classifying map. In this section we describe a procedure to determine
the regions in which the Brillouin torus can be divided so that the
restriction of $\bar{f}$ to the interior of each one of them becomes an immersion into $S^2$.
Let ${F_{1}}_{\ast}$ be the differential of $F_1$ and 
\begin{eqnarray}\label{vectors-M}
  X_{\nu}(k) 
  = {F_{1}}_{\ast}\Big(\frac{\partial}{\partial k^{\nu}}\Big)\Big|_k
  =\sum_{i}
    \frac{\partial h^{i}}{\partial k^{\nu}}\frac{\partial}{\partial x^{i}}\Big|_{F_1(k)},
\end{eqnarray}
where $\nu=\{1,2\}$, $i=\{1,2,3\}$, and the set $\{\frac{\partial}{\partial x^{i}}\}$ is the canonical basis
of $\RR^3$. 

\begin{figure}[tbp]
\centering
   \scalebox{1.11}{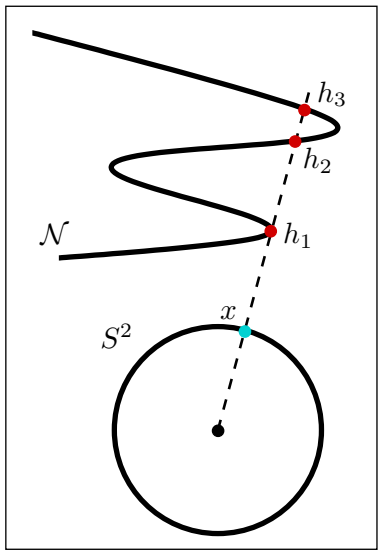}
   \caption{Color online. Schematic representation of different points $h(k)\in \mathcal{N}$ that are mapped to the same point $x\in S^2$. Note that while $h_1\in F_1(\gamma)$ -hence, the ray is tangent to $\mathcal{N}$-, $h_2$ and $h_3$ are in $F_1(\gamma')$ but not in $F_1(\gamma)$ -hence, the ray is secant to $\mathcal{N}$-.}
\label{lifting-cartoon}
\end{figure}

For each $k\in T^2$ we can form the matrix $A$ whose rows are
the vector $F_1(k)$,  $X_1(k)$ and $X_2(k)$ and consider the real function $d_\mathcal N$ defined on $T^2$  by

\begin{eqnarray}\label{det-N}
 d_{\mathcal{N}}(k) = \mbox{det}\big(A(k)\big)=\sum_{ijk}\epsilon_{ijk}
                  h^{i}(k)\frac{\partial h^{j}}{\partial k^{1}}(k)
                          \frac{\partial h^{k}}{\partial k^{2}}(k).
\end{eqnarray}
Let $\gamma$ be the solution set of the equation $d_{\mathcal{N}}(k)=0$ in $T^2$. It is given by
points where ${F_{1}}_{\ast}|_k$ does not have maximal rank, two in this case, or by
points where $F_{1}(k)$ lies in the plane spanned by $\{X_{1}(k), X_{2}(k)\}$. Similarly, $\gamma$ consists of points where the rank of $\bar{f}_{*}$ is less than $2$ (hence $\bar{f}$ is not an immersion).
The later case can be represented by the point $h_1$ in figure
(\ref{lifting-cartoon}) where $x=\bar{F}_{2}(h_1)\in S^2$. In this way, different subsets of $\mathcal{N}$ that are mapped to
the same portion of $S^2$ can be individualized. 

Note that from Eq. (\ref{berry-curvature}) and Eq. (\ref{det-N}) we have
$\Omega_{12}(k)=\frac{1}{2\vert\vert h(k) \vert\vert^{3}}~d_{\mathcal{N}}(k)$
and, therefore, the Berry curvature vanishes along $\gamma$. In addition, from
Eq. (\ref{sqrt-det-g}) we see that the metric $g$ becomes singular when
restricted to $\gamma$. The nonemptiness of $\gamma$ for two-band models when 
the parameter space is a two-torus has recently been demonstrated by Mera and Ozawa in 
reference [\onlinecite{mera}]. Here we go further exploiting this result.

Let $\tilde{\gamma}=\bar{f}(\gamma)$ and 
\begin{eqnarray}\label{lifting}
\gamma'=\bar{f}^{-1}(\tilde{\gamma}),
\end{eqnarray}
where $\bar{f}^{-1}(\tilde{\gamma})$ is the inverse image of $\tilde{\gamma}$, that is, the set of points whose image under  $\bar{f}$ lies in $\tilde{\gamma}$. 
Notice that $\gamma\subset \gamma'$, but the inclusion may be strict. That is, there may be points in $\gamma'$ that are not in $\gamma$ as we will see in the last two examples in Section \ref{results} but not in the first example, where $\gamma'=\gamma$. Notice that the quantum metric does not become singular at the points in $\gamma'$ that are not in $\gamma$. 

Graphically, this means that the ray coming from the origin and passing through $x\in\tilde{\gamma}$ may be tangent to $\mathcal{N}$ at points $h_1\in F_1(\gamma)$ or may be secant to $\mathcal{N}$ at points like $h_2$ and $h_3$ as sketched in figure~(\ref{lifting-cartoon}). We assume that
$\gamma'$ is a set of curves and that $T^2\setminus \gamma'={R}_1\cup\ldots\cup{R}_N$, where the ${R}_j$ are disjoint, open and connected sets. These $R_s$ are the connected components of $T^2\setminus \gamma'$. This assumption is satisfied in the examples analyzed in Section \ref{results}.

As a consequence, the restriction of $\bar{f}$ to each ${R}_{s}$, $\bar{f}\vert_{{R}_{s}}$,  is a local diffeomorphism from the later into $S^2$. If $\bar{f}\vert_{{R}_{s}}$ is injective, as we will assume from now on, $\mathcal{M}_{s}=\bar{f}({R}_{s})$ is a regular submanifold of $S^2$. In all the examples we will consider in Section \ref{results} this last condition is satisfied.

In addition, for each $s\in\{1,\ldots,N\}$ the restriction of $g$ to ${{R}_{s}}$  provides a  true Riemannian metric on ${R}_{s}$. In this way, the concept of a \textit{distance between points} is well defined within ${R}_{s}$. The notion of a distance turns out to be very important in different areas involving quantum geometry, such as quantum circuits.\cite{katabarwa}  

\vspace{0.5cm}
\textit{Integration over ${R}_{s}$.}
There is a well defined notion of integral over $n$-dimensional manifolds for compactly supported $n$-forms. As we have mentioned in Section
\ref{introduction}, we want to study the contributions to Eq. (\ref{cher-volume})
coming from each region ${R}_{1},\ldots,{R}_{N}$; in general,
neither of these regions fully contains the support of the corresponding
$2$-form. In Appendix \ref{integration}, we summarize a possible procedure for obtaining a well-defined integral over any ${R}_{s}$ in such a way that, for a given two-form $\omega$ on $T^2$, we obtain:

\begin{eqnarray}\label{integracion-en-Rs}
\int_{T^2}\omega&=&\sum_{s=1}^{N}\mathcal{I}_{s},\quad\text{where}\nonumber\\
\mathcal{I}_{s}&=&\int_{R_{s}}i^{\ast}_{s}\omega.
\end{eqnarray}

Regarding the integrals in Eq. (\ref{cher-volume}) we will use the notation
$\mathcal{I}_{cs}$ and $\mathcal{I}_{vs}$ to indicate the contribution of $R_{s}$  to $c$ and $v_g$ respectively,
\begin{eqnarray}\label{cher-volume-contributions}
  \mathcal{I}_{cs}&=&\frac{1}{2\pi}\int_{R_{s}}~\Omega_{12}(k)~dk^{1}\wedge dk^{2}\nonumber,
  \\
  \mathcal{I}_{vs}&=&\frac{2}{\pi}\int_{R_{s}}\sqrt{\mbox{det}(g)(k)}~dk^{1}\wedge dk^{2}.
\end{eqnarray}

We have oriented $T^2$ in such a way that the two-form $\omega_{T^2}=dk^{1}\wedge dk^{2}$ determines its orientation and as a consequence the corresponding one of $R_{s}$. Because $R_{s}$ does not intersect $\gamma$, $\Omega_{12}(k)$ of Eq. (\ref{berry-curvature}) does not vanish in this region. Therefore, $\Omega=\Omega_{12}(k)~dk^{1}\wedge dk^{2}=\Omega_{12}(k)~\omega_{T^2}$ is a nonzero two-form when restricted to $R_{s}$. Hence $\Omega_{12}(k)$ is nonzero and being continuous on the connected set $R_{s}$, it has a definite sign on $R_{s}$. As a consequence $\Omega$ and $\omega_{T^2}$ define the same (opposite) orientation in $R_{s}$ if $\Omega_{12}(k)>0$ ($\Omega_{12}(k)<0$).

On the other hand, since $S^2$ is orientable, there are two-form on $S^2$ which do not vanish at any point. In particular, since the Berry curvature of the tautological line bundle $\Theta$ is given by a nonzero two-form, $\tilde{\Omega}$, on the base $\CC\PP^1\simeq S^2$, it can be chosen to define the orientation. In APPENDIX \ref{pullback} we show that it defines an orientation of $S^2$ equivalent to the one given by its exterior normal when it is embedded in $\RR^3$. Also APPENDIX \ref{pullback} shows that $\Omega=\bar{f}^{\ast}\tilde{\Omega}$, that is, the former is a pullback construction from the latter. That is, the Berry curvature on the torus is the pullback of the Berry curvature on the tautological line bundle. By definition, the diffeomorphism $\bar{f}: R_{s}\rightarrow \mathcal{M}_{s}=\bar{f}(R_{s})$ is orientation preserving if $\Omega=\bar{f}^{\ast}\tilde{\Omega}=\lambda\omega_{T^2}$, where $\lambda$ is a positive real function. As a consequence, our procedure for splitting $T^2$ provides a relation between the positive (negative) sign of the Berry component $\Omega_{12}$ in $R_{s}$ with the orientation-preserving (reversing) behavior of the classifying map $\bar{f}$. We will exploit this relation when analyzing the examples in the following section.

In addition, by virtue of Eq. (\ref{sqrt-det-g}), $\pm\sqrt{\mbox{det}(g)}= \Omega_{12}/2$, and Eq. (\ref{cher-volume-contributions}) we obtain the following remarkable first result  
\begin{equation}\label{local-saturated-inequality}
 \mathcal{I}_{vs}=\pm 2\mathcal{I}_{cs},
\end{equation}
where the positive (negative) sign can now be traced back to the orientation-preserving (reversing) character of
the diffeomorphism $\bar{f}|_{R_{s}}$.
Furthermore, if $R_{s}$ and $R_{s'}$ are two of the open sets such that
\begin{eqnarray}\label{equal-integrals}
  \bar{f}(R_{s})=\bar{f}(R_{s'}) \stext{ then }
\begin{cases}
    \mathcal{I}_{cs}=\pm\mathcal{I}_{cs'},\\
    \mathcal{I}_{vs}=~~\mathcal{I}_{vs'},
\end{cases}
\end{eqnarray}
where the sign is determined by the preserving/reversing orientation behavior of $\bar{f}$ on both $R_s$ and $R_{s'}$.

Note that Eq. (\ref{local-saturated-inequality}) is the local (always saturated) version of the relation $v_g\ge2\vert c\vert$ found by Roy in ref. [\onlinecite{roy}] and by Mera and Ozawa in ref. [\onlinecite{mera}]. Note, in addition, that while the Chern number, $c=\sum_{s=1}^{N}\mathcal{I}_{cs}$ is guaranteed to be an integer, this is not in general the case for each contribution $\mathcal{I}_{cs}$.

As an application of the previous construction, we can find subsets of $T^2$ whose volume according to $g$ is $v_{min}$. We may procceed as follows. If $c=1$ ($c=-1$), then we select all the individual regions where $\Omega_{12}>0$ ($\Omega_{12}<0$). Next, within this subset, we classify the regions according to the multiplicity of their image under $\bar{f}$. This means that if there are $n$ regions that are mapped onto the same subset $\mathcal{M}_s \subset S^2$, then the multiplicity is $n$. Now, for those with $n>1$, we select one of them and group these with all the other regions with $n=1$. The result is a subset of $T^2$ with volume $v_{min}$ and where the integral of $\Omega_{12}/2\pi$ is $c$. The same process can be
straightforwardly generalized to the case where $|c|\ge2$.
Regarding the excess of volume, $v_{ex}$, note that the regions that were not used to form a minimal volume region, do not contribute to $c$. Moreover, those regions arising from multiplicity $n>1$ that were not included in
the region of minimal volume should be mapped by $\bar{f}$ onto the same subset of $S^2$, but with opposite orientation compared to the ones with $\Omega_{12}<0$ ($\Omega_{12}>0$). In this way, they contribute to $v_{ex}$ but not to $c$.

As a final remark of this section, we want to comment on another important concept related to the Chern number, that is the degree of a smooth map. Let $M$ and $N$ be compact, connected, oriented, and smooth manifolds of the same dimension $n$ and let $F:M \rightarrow N$ be a smooth map. Then, the degree of $F$ is the unique number $\delta$ that satisfies
\begin{equation}\label{degree}
 \int_{M}F^{\ast}\omega = \delta\int_{N}\omega ~~,
\end{equation}
for every smooth $n$-form on $N$; it happens to be an integer, [\onlinecite{lee-book},\S17]. In the case of $N=S^2$, $M=T^2$, $\omega=\tilde{\Omega}$, and $F=\bar{f}$ then from Eq. (\ref{cher-volume}) the LHS of Eq. (\ref{degree}) is equal to $2\pi c$, while the RHS of Eq. (\ref{degree}) is equal to $2\pi \delta$. Therefore, $c=\delta$. It is known that homotopic maps have the same degree, so we obtain the same result for both classifying maps, $\bar{f}$ and ${f}$. The degree $\delta$ has a simple geometrical meaning, that is how many times and with which orientation $S^2$ is fully covered by $T^2$ via the map $\bar{f}$. There is also a characterization of the degree of a map counting (with a sign related to orientation preservation or not) the elements in the preimage of a regular value of the map. This notion will help with the interpretation of the following examples.


\section{Results}\label{results}

Here we analyze three examples of topological insulators in two dimensions by using the definitions and results of Sections \ref{2band-models} and \ref{splitting} . The first one corresponds to a time-reversal Hamiltonian modeling the experiment of reference [\onlinecite{tan}]. Following that, we will focus on two different models for Chern insulators, the simplest model with nearest-neighbor hopping on a square lattice and the Haldane model defined on a honeycomb lattice. 

\subsection{A time-reversal model}\label{TR-model-section}

As mentioned in the introduction, the recent experiment by Tan \textit{et al.}, see reference [\onlinecite{tan}], reported different methods to directly measure the quantum metric tensor and explored a topological phase transition in a simulated time-reversal two-band system. Since the Chern number vanishes because of time-reversal symmetry, the phases were classified by the so called Euler characteristic number of the occupied band.

Here we study the corresponding model making use of the procedure presented in the previous section. To begin with, the model is given by $F_1: T^2 \rightarrow \RR^3$ such that $F_1(k)=(h^{1}(k), h^{2}(k), h^{3}(k))$ for
\begin{eqnarray}\label{TR-model}
 h^{1}(k) &=& \mbox{sin}(k^{1})\mbox{sin}(k^{2})\nonumber\\
 h^{2}(k) &=& \mbox{sin}(k^{1})\mbox{cos}(k^{2})\\
 h^{3}(k) &=& m+\mbox{cos}(k^{1})\nonumber.
\end{eqnarray}
The model describes spinless free fermions derived from the many-body Hamiltonian
of the XY spin chain after the Jordan-Wigner transformation. \cite{yu-quan-ma}

Using Eq. (\ref{TR-model}) and then Eqs. (\ref{det-N}), (\ref{berry-curvature}) and (\ref{sqrt-det-g}) the expressions for $d_{\mathcal{N}}(k)$, the Berry curvature and the quantum volume for this model are given by
\begin{eqnarray}\label{expressions-tr}
 d_{\mathcal{N}}(k) &=& \mbox{sin}(k^{1})\big(m~\mbox{cos}(k^{1})+1\big)\nonumber\\
 \Omega_{12}(k)&=& \frac{\mbox{sin}(k^{1})\big(m~\mbox{cos}(k^{1})+1\big)}
 {2\big(1 + m^2 + 2m~\mbox{cos}(k^{1})\big)^{3/2}} \\
 \sqrt{\mbox{det}g(k)}&=& \frac{\vert \mbox{sin}(k^{1})\big(m~\mbox{cos}(k^{1})+1\big) \vert}{4\big(1 + m^2 + 2m~\mbox{cos}(k^{1})\big)^{3/2}} \nonumber.
\end{eqnarray}

Note that the set $\mathcal{N}=F_1(T^2)$ is a radius one two sphere, $S^2$, centered at $(0, 0, m)$. For $\vert m \vert > 1$, the origin of coordinates is outside $\mathcal{N}$ and then $\mathcal{M}=\bar{F}_2(\mathcal{N})$ is not all of $S^2$. Thinking of $c$ as the degree of the map $\bar{f}$, which is not onto, this means $c=0$, suggesting a trivial phase of the model. On the other hand, as soon as $\vert m \vert < 1$ the origin is contained within the volume enclosed by $\mathcal{N}$ and $ \mathcal{M}=S^2$, opening the possibility for a non-trivial phase and, therefore, a TPT at $\vert m \vert=1$. However, due to the constraint imposed by the time-reversal symmetry, $\Omega_{12}(-k)=-\Omega_{12}(k)$, the Chern number remains $c=0$ so we cannot classify the phases as trivial and non-trivial in the topological sense. As a consequence, there is no obstruction to globally defining the eigenstates of the Bloch Hamiltonian over the Brillouin torus, i.e., the Bloch valence bundle $\mathcal{V}$ is trivial. Taking the case $m=0$ as an example, although not physically relevant,  
$|\psi(k)\rangle=\big( i~\mbox{sin}(k^1/2)e^{ik^2}, \mbox{cos}(k^1/2)\big)^{T}$ is a globally defined and nowhere vanishing section of the valence eigenspace.

Note that as $k^1\in [-\pi,\pi]$, the angle defined between $F_1(k)$ and the $h^3$ axis, runs twice the usual polar angle of the spherical parametrization of $S^2$ centered at $(0, 0, m)$. Therefore when $\vert m \vert < 1$, $\mathcal{N}$ covers twice the unit sphere centered at the origin independently of $m$. Here we will show that the coverings have opposite orientations leading to $c=0$. Note that these 'coverings' are not covering spaces in the mathematical sense. 

\begin{figure}[tbp]
\centering
   \scalebox{1.11}{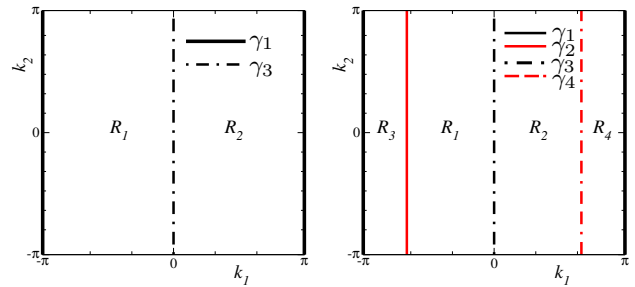}
   \caption{Color online. Different regions in which the torus is divided by the curves $\gamma_\alpha$ for the selected values $|m|<1$ (left) and $m=2$ (right).  }
\label{tr_regions}
\end{figure}

The solution of the equation $d_{\mathcal{N}}(k)=0$ is given by the set of curves $k^1=0, \pm\pi$ (where $k^1=-\pi$ and $k^1=\pi$ are the same curves in $T^2$) when $\vert m \vert <1$ with the additional curves $k^1=\pm \mbox{Arccos}(-1/m)$ when $\vert m \vert >1$. There are no other solutions coming from Eq. (\ref{lifting}). The curves as well as the regions in which $T^2$ is divided are shown in figure (\ref{tr_regions}) for two selected values of $m$. 

We obtain that the total number of regions $N_{T^2}=\sum_{s=1}^{N}1$ remains constant at the value $N_{T^2}=2$ ($N_{T^2}=4$) for all $\vert m \vert <1$ ($\vert m \vert >1$). So, $N_{T^2}$ is a topological invariant for the family of Hamiltonians defined by $F_1(k)$ and smoothly deformed by $m$. Then, we are able to distinguish two different regimes of the model. Equivalently they can be distinguished by the location of the origin of $\RR^3$, inside or outside the set $\mathcal{N}$ and, also, by the surjectivity of the map $\bar{f}$. 

\begin{figure}[tbp]
\includegraphics[scale=.50]{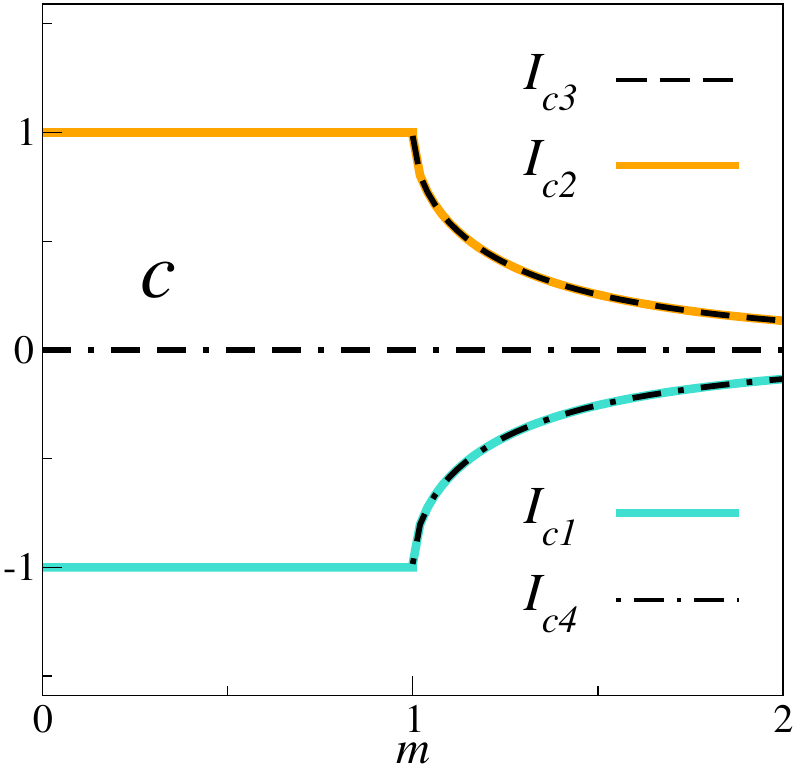}
\includegraphics[scale=.50]{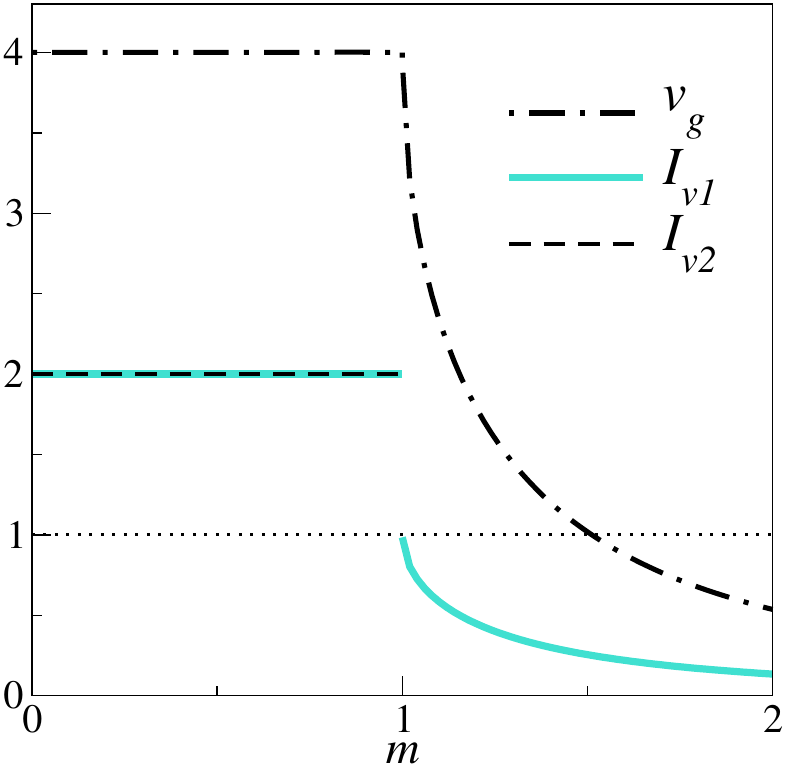}
\caption{ Color online. Top panel: Chern number, $c$ (dashed-dashed-dot line), and its contributions coming from the integrals ${\mathcal{I}_c}_{s}$ as functions of the parameter $m$ for the time-reversal system. Bottom panel: quantum volume, $v_g$, and its contributions coming from the integrals ${\mathcal{I}_v}_{s}$. For $m>1$ only ${\mathcal{I}_v}_{1}$ is displayed and ${\mathcal{I}_v}_{1}={\mathcal{I}_v}_{2}={\mathcal{I}_v}_{3}={\mathcal{I}_v}_{4}$.}
\label{deg_vol_tr}
\end{figure}

Now we discuss some features of the $\vert m \vert <1$ regime. To begin with the splitting procedure of Section \ref{splitting}, removing $\gamma_1$ and $\gamma_3$ in the left panel of figure (\ref{tr_regions}) splits $T^2$ into two different regions, ${R}_{1}$ and ${R}_{2}$. The restriction of $\bar{f}$ to each one of them being a parametrization of the two-sphere (up to the poles). Note that the fact that Eq. (\ref{lifting}) does not have other solutions than the ones in which $d_{\mathcal{N}}(k)=\Omega_{12}(k)=0$, becomes relevant for considering ${R}_{1}$ and ${R}_{2}$ parametrizations of $S^2$ (that is $\bar{f}$ restricted to them is a diffeomorphism), otherwise injectivity of $\bar{f}$ is not guaranteed.  This is to say that $\bar{f}$ maps diffeomorphically each one of ${R}_{1}$ and ${R}_{2}$ onto $S^2\setminus {L}$ with ${L}$ a single meridian from the north to south pole. However, while $\Omega_{12}(k)<0$ for all $k\in {R}_{1}$, $\Omega_{12}(k)>0$ for all $k\in {R}_{2}$, so the diffeomorphism is orientation-reversing in ${R}_{1}$ and orientation-preserving in ${R}_{2}$.\cite{orientation}
As a consequence, from Eq. (\ref{cher-volume-contributions}) we obtain $-{\mathcal{I}_c}_{1}={\mathcal{I}_c}_{2}=1$ being this behavior independent of the value of $m$.  Therefore, in this regime $c={\mathcal{I}_c}_{1}+{\mathcal{I}_c}_{2}=0$, as it is shown in the top panel of figure (\ref{deg_vol_tr}). 
Regarding the contributions to the quantum volume, from Eq. (\ref{local-saturated-inequality}) we obtain ${\mathcal{I}_v}_{1}={\mathcal{I}_v}_{2}=2$. We have also verified this result by direct computation of Eq. (\ref{cher-volume-contributions}), see bottom panel of figure (\ref{deg_vol_tr}).

On the other hand, when $\vert m \vert >1$, the four regions in the right panel of figure (\ref{tr_regions}) are mapped onto the same subset of $S^2$, a beanie centered at the north pole and it is topologically equivalent to an open disk. That is, as a set of points we obtain $\mathcal{M}_{1}=\mathcal{M}_{2}=\mathcal{M}_{3}=\mathcal{M}_{4}=\mathcal{M} \varsubsetneq S^2$. We want to emphasize that $\mathcal{M}$ is not $S^2$ while in the case $\vert m \vert <1$, $\mathcal{M}=S^2$. Notice that $\bar{f}$, while injective, behaves differently on each $\mathcal{M}_{j}$: it is an orientation-preserving diffeomorphism with its image when restricted to $R_2$ and $R_3$, it is orientation-reversing on $R_1$ and $R_4$. Accordingly we obtain ${\mathcal{I}_c}_{2}={\mathcal{I}_c}_{3}=-{\mathcal{I}_c}_{1}=-{\mathcal{I}_c}_{4}>0$. The top panel of figure (\ref{deg_vol_tr}) shows the numerical evaluations of ${\mathcal{I}_c}_{s}$ for $s=1,2,3$ and $4$ as a function of $m>0$. Reversing the sign of $m$ does not introduce any significant change (when $m<0$, letting $m\rightarrow0$, makes the center of $\mathcal{N}$ approach the origin from below). Note that $c=\sum_{s=1}^{N}\mathcal{I}_{cs}=0$ as expected.
However, the contributions to $v_g$ are such that ${\mathcal{I}_v}_{1}={\mathcal{I}_v}_{2}={\mathcal{I}_v}_{3}={\mathcal{I}_v}_{4}$. The bottom panel of figure (\ref{deg_vol_tr}) shows the results for $v_g$ as well as ${\mathcal{I}_v}_{s}$. For $m>1$ only one representative of the equal  ${\mathcal{I}_v}_{s}$ for $s=1,2,3$ and $4$ is shown.  

Our final comment of this section is regarding the interpretation of $v_g$ as the surface contribution to the Euler characteristic number, $\chi$, of $\mathcal{M}$ that has been provided in references \onlinecite{tan, zhu, tan-erratum}. From topology $\chi({T^2})=0$ and $\chi(\mathcal{M})=2$ when $\mathcal{M}=S^2$ ($\vert m \vert <1$) or $\chi(\mathcal{M})=1$ when $\mathcal{M}\varsubsetneq S^2$ ($\vert m \vert >1$). There is no other relevant set of points underlying this model.
We recall that for a closed, oriented two-dimensional manifold ${X}$ endowed with a Riemannian metric, the Gauss-Bonnet Theorem relates $\chi$ to the Gaussian curvature, $\mathcal{K}$,
\begin{equation}\label{gauus-bonnet}
 \chi({X})=\frac{1}{2\pi}\int_{{X}} \mathcal{K}~dA
\end{equation}
where $dA$ is the area element. For instance if $X=S^2(1/2)$ (the sphere of radius $1/2$) $\mathcal{K}=4$ and the right hand side (RHS) of Eq. (\ref{gauus-bonnet}) is $2$. Alternatively, the two-sphere can be locally parameterized, in the sense of Appendix \ref{integration}, by the open sets $R_s\subset T^2$ ($s=\{1,2\}$, see the left panel of figure (\ref{tr_regions})) by using $\bar{f}$ Eq. (\ref{gauus-bonnet}) can be rewritten as
\begin{eqnarray}\label{pullback-gauus-bonnet}
\chi({S^2}) &=&\frac{(-1)^{f_{s}}}{2\pi}\int_{{R_s}} \big(\bar{f}\vert_{{R}_{s}}\big)^{\ast}\big( \mathcal{K}~dA\big)\nonumber\\
&=&\frac{(-1)^{f_{s}}}{2\pi}\int_{{R_s}} \Big(\big(\bar{f}\vert_{{R}_{s}}\big)^{\ast}\mathcal{K}\Big)\Big(\big(\bar{f}\vert_{{R}_{s}}\big)^{\ast}dA\Big)\\
&=&\frac{(-1)^{f_{s}}}{2\pi}\Big(\big(\bar{f}\vert_{{R}_{s}}\big)^{\ast}\mathcal{K}\Big)\int_{{R_s}}\big(\bar{f}\vert_{{R}_{s}}\big)^{\ast}dA  \nonumber\\
&=&\frac{(-1)^{f_{s}}}{2\pi}4\int_{{R_s}}  (-1)^{f_{s}}\sqrt{\mbox{det}(g)(k)}~dk^{1}dk^{2}\nonumber\\
&=&\frac{2}{\pi}\int_{{R_s}}  \sqrt{\mbox{det}(g)(k)}~dk^{1}dk^{2}\nonumber\\
&=& v_{g}/2\nonumber,
\end{eqnarray}
where we have used the definition of $v_g$ introduced in Eq. (\ref{cher-volume}) and the fact that $(-1)^{f_{s}}=+1$(-1) if the diffeomorphism $\bar{f}\vert_{{R}_{s}}$ preserves(reverses) the orientation between ${R_s}$ and $S^2$. Furthermore, the pullback of a constant results in the same constant $\big( \bar{f}\vert_{{R}_{s}}\big)^{\ast}\mathcal{K}=4$ and the pullback of $dA$ is, up to the sign $(-1)^{f_{s}}$, the volume element associated to the quantum metric.

This example shows why $v_g=2\chi({S^2})=4$ when $\vert m \vert <1$. We want to stress that the use of Eq. (\ref{lifting}) together with the fact that $\bar{f}|_{R_s}$ being injective as well as $\bar{f}(R_s)=S^2$ (up to zero measure subset) are key for the appearence of the multiplicity factor $2$ in the previous relation.
Similarly, when $\vert m \vert >1$ the beanie $X=\mathcal{M}$ is a manifold with smooth boundary and it can be locally parameterized by $R_s\subset T^2$ ($s\in\{1,2,3,4\}$, see the right panel of figure (\ref{tr_regions})). The analog of Eq. (\ref{gauus-bonnet}) valid for manifolds with smooth boundary adds to its right hand side a contribution coming from the boundary of ${X}$, $I_{\partial X}$. A computation similar to Eq. (\ref{pullback-gauus-bonnet})  gives $v_g/4+I_{\partial \mathcal{M}}$.    

From the two cases discussed above, it is clear that $v_g$ does not match with the surface contribution to $\chi(\mathcal{M})$ since it carries the multiplicity of $\bar{f}$, that is, the number of regions in $T^2$ that are mapped to the same portion of $S^2$.
Therefore, we do not find an appropriate framework in which the concept of the Euler characteristic number can be used. Instead, we agree with Mera and Ozawa in references \onlinecite{mera, ozawa} where $v_g$ is treated as the volume of $T^2$ when the quantum metric is a Riemannian metric and as a simple notation for the integral in Eq. (\ref{cher-volume}) when it is not.  
\subsection{Square model}\label{simplest-model-section}

\begin{figure}[tbp]
  \centering
  \scalebox{1.0}{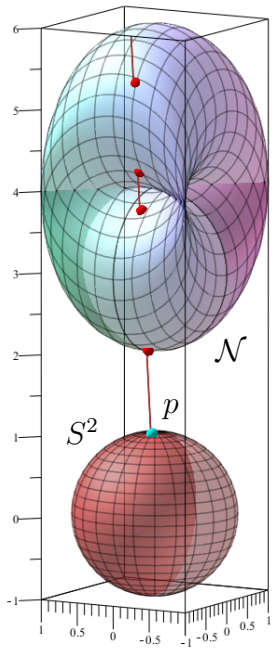}
  \caption{Color online. Subset $\mathcal{N}$ for the square model of Section \ref{simplest-model-section} for $m=4$, $S^2$ and an example of a ray showing the set of different points in $\mathcal{N}$ that are mapped to the same point near the north pole in $S^2$. }
  \label{N_and_S2_with_intersection_points-6}
\end{figure}

The simplest model for a Chern insulator is given by the following real functions 
\begin{eqnarray}\label{simplest-model}
 h^{1}(k) &=& \mbox{sin}(k^{1})\nonumber\\
 h^{2}(k) &=& \mbox{sin}(k^{2})\\
 h^{3}(k) &=& m + \mbox{cos}(k^{1}) + \mbox{cos}(k^{2})\nonumber,
\end{eqnarray}
where $k\in [-\pi,\pi]^2$.
The model can be seen as a part of a larger one considered by Qi \textit{et al.} \cite{qi} and
realized in Hg$_{1-x}$Mn$_{x}$Te/Cd$_{1-x}$Mn$_{x}$Te quantum wells by Liu \textit{et al.} \cite{liu}. 
The parameter $m$ in the model is responsible for displacements of $\mathcal{N}$ in the $x^{3}$ direction of $\RR^3$ and therefore can be used to drive the system into a TPT. In figure (\ref{N_and_S2_with_intersection_points-6}) we show $\mathcal{N}$ for $m=4$ together with $S^2$ and the set of points $\bar{F}_{2}^{-1}(p)\subset\mathcal{N}$ where $p\in S^2$ is a point near the north pole.

From Eq. (\ref{simplest-model}) we derive the following expressions, 
\begin{eqnarray}\label{expressions-simplest}
 d_{\mathcal{N}}(k) &=& m~\mbox{cos}(k^{1})\mbox{cos}(k^{2})+\mbox{cos}(k^{1})+\mbox{cos}(k^{2})\nonumber\\
 \Omega_{12}(k)&=& \frac{d_{\mathcal{N}}(k)}
 {2\vert \vert h(k)\vert \vert^{3}} \\
 \sqrt{\mbox{det}(g(k))}&=& \frac{\vert \Omega_{12}(k) \vert}{2} \nonumber
\end{eqnarray}

\begin{figure}[tbp]
\centering
\includegraphics[scale=.50]{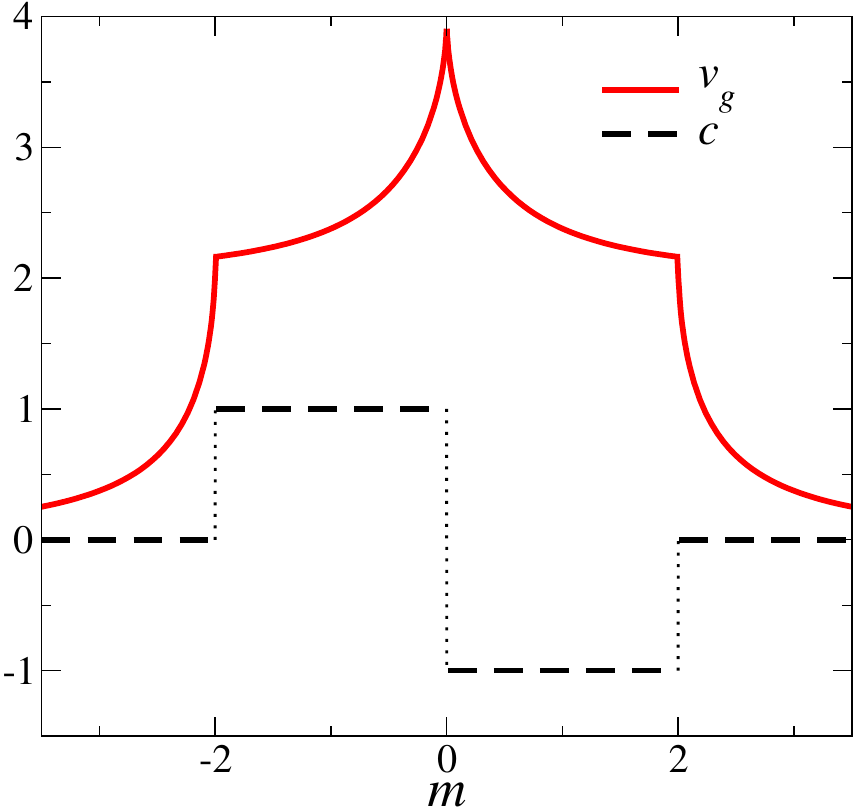}
\caption{ Color online. Chern number, $c$, and the quantum volume, $v_g$, as functions of $m$ for the square model. A topologically trivial (non-trivial) phase corresponds to $\vert m\vert >2$ ($\vert m\vert<2$). }
\label{deg_vol_simplest}
\end{figure}

The Chern number, $c$, and the quantum volume, $v_g$, as functions of $m$ are shown in figure (\ref{deg_vol_simplest}). Both magnitudes were first calculated in reference \onlinecite{yang-2015}.  

According to the value of $c$, topologically trivial and non-trivial phases correspond to $\vert m\vert>2$ and $\vert m\vert<2$ respectively. 

\begin{figure}[tbp]
\centering
   \scalebox{1.0}{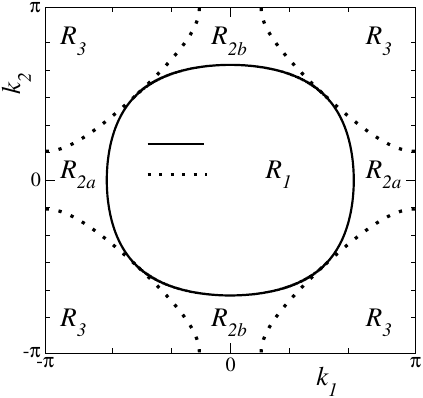}
   \scalebox{1.0}{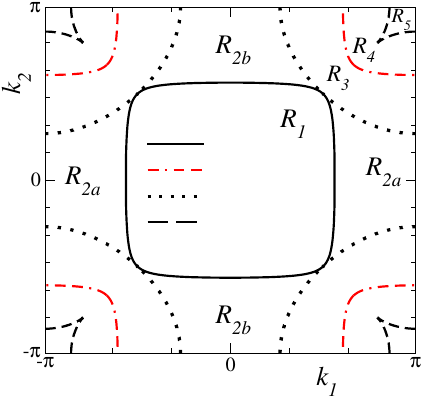}
   \caption{Color online. Different regions in which the torus is divided by the curves $\gamma_\alpha$ for the selected values $m=1$ and $m=4$ representing a topologically non-trivial (top) and trivial (bottom) phase of the square model respectively.}
\label{simplest_regions}
\end{figure}

Regarding the division of $T^2$ using the splitting procedure of Section \ref{splitting}, here we discuss the solution of $d_{\mathcal{N}}(k)=0$ in both phases. In the case $\vert m\vert<2$ the solution is given by the closed curve $\gamma_1$, $k^2=\pm \mbox{Arccos}\big(\frac{-\mbox{cos}(k^{1})}{m~\mbox{cos}(k^{1})+1}\big)$, centered at the origin and plotted as a solid line in the top panel of figure (\ref{simplest_regions}). The solution of Eq. (\ref{lifting}) for $\tilde{\gamma}_1=\bar{f}({\gamma}_1)$ has two components: ${\gamma}_1$ and another closed curve, which we call $\gamma_3$, centered at the corner $(\pi,\pi)$ (identified with the other three corners of $T^2$ in figure (\ref{simplest_regions})) and it is represented by the black dotted line in figure (\ref{simplest_regions}).

When $\vert m \vert > 2$, the solution of $d_{\mathcal{N}}(k)=0$ is composed of $\gamma_1$ and an additional closed curve $\gamma_2$ centered at the corner $(\pi,\pi)$ and plotted for $m=4$ as a dot-dashed red line in the bottom panel of figure (\ref{simplest_regions}). Regarding Eq. (\ref{lifting}) for $\gamma_1$, in this case the solution contains an additional component $\gamma_4$ centered at the corner $(\pi,\pi)$ shown for $m=4$ in the lower panel of figure (\ref{simplest_regions}).
On the other hand, the solution of Eq. (\ref{lifting}) for $\tilde{\gamma}_2=\bar{f}({\gamma}_2)$ is the curve ${\gamma}_2$ itself.
The name we choose for each open set in which the curves divide $T^2$ is indicated in figure (\ref{simplest_regions}). 

\begin{figure}[tbp]
\includegraphics[scale=.50]{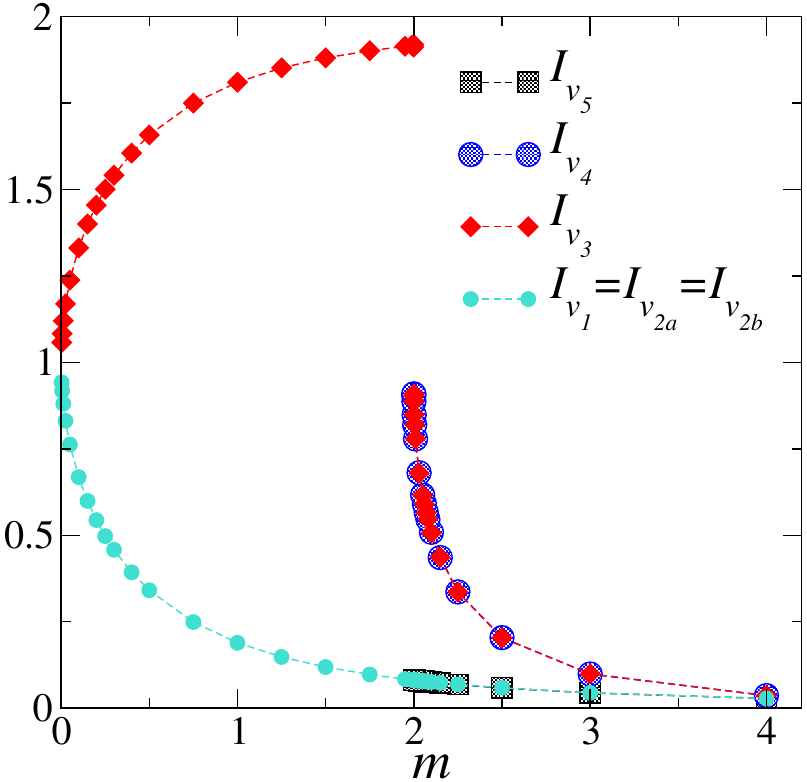}
\includegraphics[scale=.50]{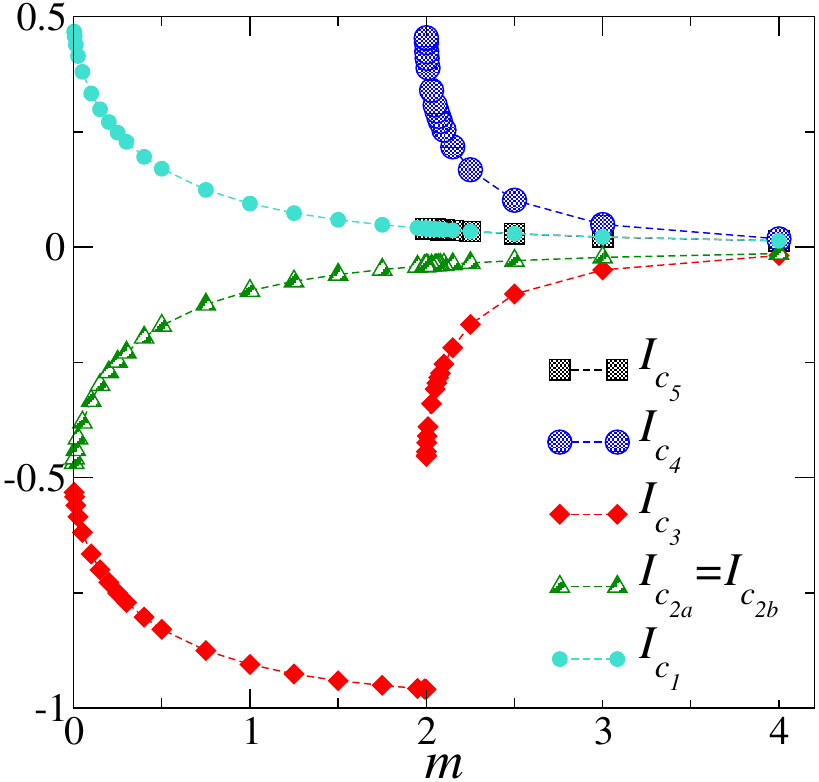}
\caption{Color online. Partial contributions to the quantum volume, ${\mathcal{I}_v}_{s}$, (top panel) and Chern number, ${\mathcal{I}_c}_{s}$, (bottom panel) as functions of $m$ for the square model.}
\label{vol-vs-m}
\end{figure}

Now we turn the discussion to the analysis of ${\mathcal{I}_v}_{s}$ and ${\mathcal{I}_c}_{s}$. Their values as functions of $m$ are shown in the top and bottom panels of figure (\ref{vol-vs-m}) respectively. Note that $v_g = \sum_{s}{\mathcal{I}_v}_{s}$ and $c=\sum_{s}{\mathcal{I}_c}_{s}$. The sign of the integrals ${\mathcal{I}_c}_{s}$ reflects the orientation-preserving (positive sign) or orientation-reversing (negative sign) character of the classifying map $\bar{f}$ when restricted to each region ${R}_{s}$. 

\begin{figure}[tbp]
\centering
   \scalebox{0.95}{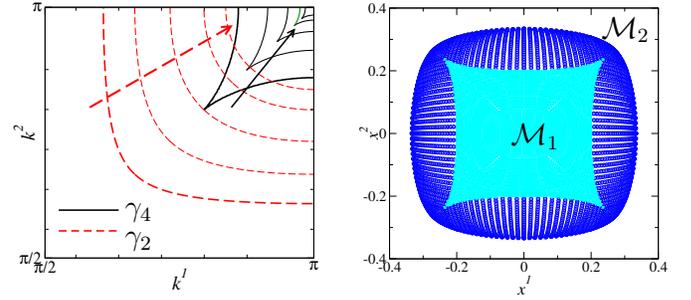}
   \caption{Color online. Left: evolution of the curves $\gamma_2$ and $\gamma_4$ as  $m$ decreases from $m=4$ to $m=2.1$. The arrows indicate the decreasing direction of $m$. Right: projection of $\mathcal{M}_1$( inner cyan pattern) and $\mathcal{M}_{2}$ (exterior blue pattern) into the plane $x^1x^2$ for $m=4$. }
\label{corner-cunia-vs-m}
\end{figure}

\begin{figure}[tbp]
\centering
   \scalebox{.75}{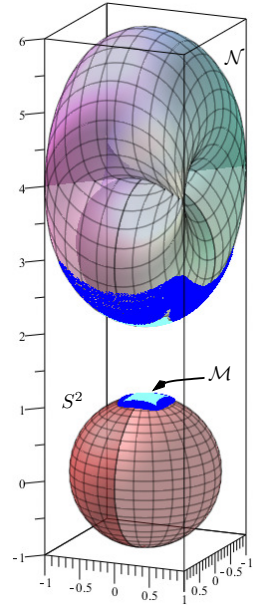}
   \caption{Color online. $F_{1}({R}_{4})\subset\mathcal{N}$ and $\bar{f}({R}_{4})=\mathcal{M}_{1}\subset S^2$ in cyan. $F_{1}({R}_{5})\subset\mathcal{N}$ and $\bar{f}({R}_{5})=\mathcal{M}_{2}\subset S^2$ in blue. $\mathcal{M}=\overline{\mathcal{M}_{1}}\bigcup\overline{\mathcal{M}_{2}}$. $m=4$. }
\label{fig:N_M_and_S2-color-10}
\end{figure}

Starting with the trivial phase, $\vert m \vert > 2$, the regions $\overline{{R}_{1}}$, $\overline{{R}_{2a}}$, $\overline{{R}_{2b}}$ and $\overline{{R}_{5}}$ are mapped onto the same square shaped subset $\overline{\mathcal{M}_{1}}\subset\mathcal{M} \subset S^2$ (which is contractible) centered at the north pole of $S^2$ (see figure (\ref{fig:N_M_and_S2-color-10})). The right panel of figure (\ref{corner-cunia-vs-m}) shows the projection of $\mathcal{M}_{1}$ (gray pattern) into the plane perpendicular to the $x^3$ axis for $m=4$. The data in the bottom panel of figure (\ref{vol-vs-m}) verifies the relation ${\mathcal{I}_c}_{1}={\mathcal{I}_c}_{5}=-{\mathcal{I}_c}_{2a}=-{\mathcal{I}_c}_{2b}$ in agreement with Eq. (\ref{equal-integrals}).

Similarly, the regions ${R}_{3}$ and ${R}_{4}$ are mapped onto the same ring shaped subset $\mathcal{M}_{2}\subset\mathcal{M}\subset S^2$ (non-contractible) with a hole in its center identical to the set $\overline{\mathcal{M}_{1}}$. This is because $\bar{f}$ (restricted to the $R_s$) is a diffeomorphism and $R_s$ having the topology of a cylinder. As a consequence of the different orientation behavior of $\bar{f}$ when restricted to each region ${\mathcal{I}_c}_{3}=-{\mathcal{I}_c}_{4}$, see figure (\ref{vol-vs-m}). The sizes of $\mathcal{M}_{1}$ and of $\mathcal{M}_{2}$ decrease as $m$ increases due to the displacement of $\mathcal{N}$ along the $x^3$ axis and therefore all the integrals ${\mathcal{I}_v}_{s}$ and ${\mathcal{I}_c}_{s}$ go to zero in the limit $\vert m \vert \rightarrow \infty$. The previous analysis explains in detail the underlying structure of the value $c = 0$ in the trivial phases. Note that $\mathcal{M}=\overline{\mathcal{M}_{1}}\bigcup\overline{\mathcal{M}_{2}}$.

A feature of the curves $\gamma_2$ and $\gamma_4$ is that they are not present for $\vert m \vert < 2$, that is, in the topologically non trivial phases. In fact, both curves shrink to the corners $(\pi,\pi)$ as $\vert m \vert \rightarrow 2$ from above. This behavior is shown in the left panel of figure (\ref{corner-cunia-vs-m}). As a consequence, the union ${R}_{4}\bigcup{R}_{5}$ is continuously shrinking and disappears. However we find that while $\vert m \vert$ decreases ${\mathcal{I}_v}_{4}+{\mathcal{I}_v}_{5}$ increases and goes to $1$. From a geometrical point of view this can be understood as follows. Recall that there are several regions of $T^2$ that are mapped into $\mathcal{M}$. The one coming from ${R}_{4}\bigcup{R}_{5}$ corresponds to the lowest portion of the set $\mathcal{N}$ facing towards the northern hemisphere of $S^2$, see Figure (\ref{fig:N_M_and_S2-color-10}). Its projection to $S^2$, namely $\mathcal{M}$, increases its size as $\vert m \vert$ decreases due to the movement of $\mathcal{N}$ towards the origin reaching its maximum at that point. Finally, recall that ${\mathcal{I}_v}_{4}+{\mathcal{I}_v}_{5}$ is nothing but a measure of $\mathcal{M}$'s volume which is equal to the quantum volume of ${R}_{4}\bigcup{R}_{5}$. On the other hand, from a physical point of view, we note that the Berry curvature increases its weight around the corner while $\vert m \vert \rightarrow 2$ from above. In fact, the shape of $\gamma_2$ tends to a circle of radius $r=\sqrt{2\epsilon}$, for $\epsilon=m-2$ and $0<\epsilon\ll1$ and the Berry curvature inside ${R}_{4}\bigcup{R}_{5}$ approaches to $\Omega_{12}(k)\sim \epsilon/[2(\epsilon^2+\vert k\vert^2)^{3/2}]$. This is a well known result from the Dirac-like behavior of the model near the Dirac points, the corners in the present case.
By using this fact, the integrals within the considered region are given by ${\mathcal{I}_v}_{4}+{\mathcal{I}_v}_{5}\sim(1-\sqrt{\epsilon/2})$ and ${\mathcal{I}_c}_{4}+{\mathcal{I}_c}_{5}\sim(1/2-\sqrt{\epsilon/8})$ in agreement with the calculated values in figure (\ref{vol-vs-m}).

\begin{figure}[tbp]
\centering
   \scalebox{1.0}{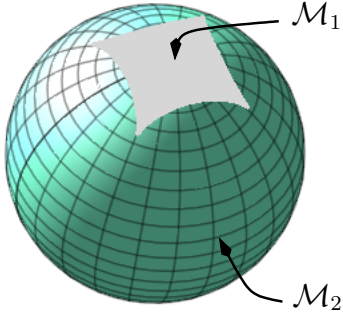}
   \caption{Color online. $\mathcal{M}_{1}$ and $\mathcal{M}_{2}$ for $m=1.99$.}
\label{M_1_and_M2_m=1.99}
\end{figure}

\begin{figure}[tbp]
\centering
\includegraphics[scale=.50]{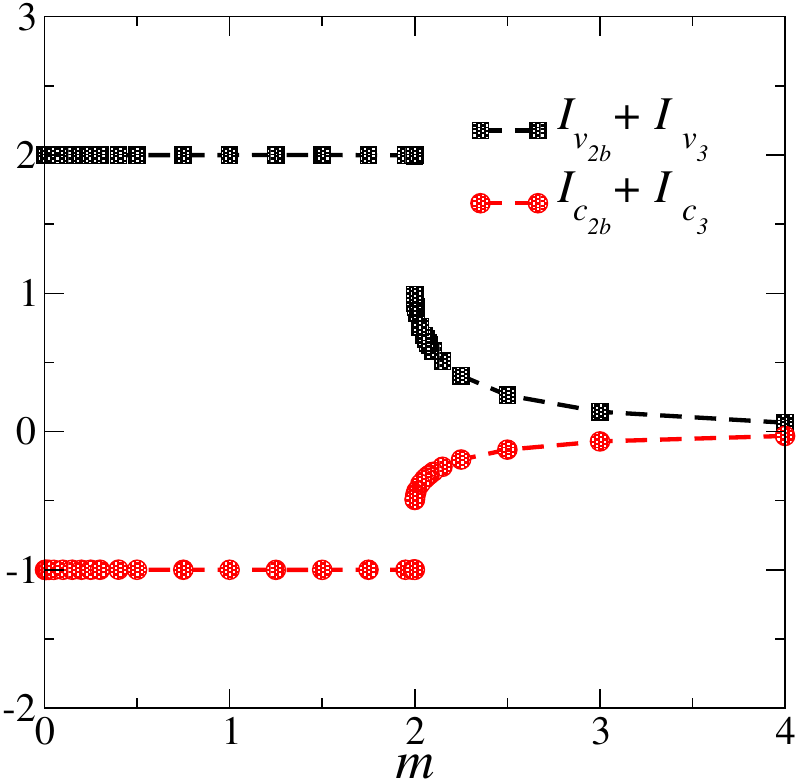}
\caption{Color online. Integrals of Eq. (\ref{cher-volume-contributions}) on the region ${R}_{2b}\bigcup{R}_{3}$ as functions of $m$.}
\label{v_2a3_vs_m}
\end{figure}

On the other hand, in the topologically non trivial phase, the regions ${R}_{1}$, ${R}_{2a}$ and ${R}_{2b}$ are mapped onto the same square shaped surface $\mathcal{M}_{1}$ as before. We obtain ${\mathcal{I}_c}_{2a}={\mathcal{I}_c}_{2b}=-{\mathcal{I}_c}_{1}$ and ${\mathcal{I}_v}_{2a}={\mathcal{I}_v}_{2b}={\mathcal{I}_v}_{1}$ in accordance with Eq. (\ref{equal-integrals}). Note that these integrals are continuous functions of $m$, that is, they do not jump at the TPT. 

Remarkably, $\overline{\mathcal{M}_{2}}=\bar{f}(\overline{R}_{3})$ is now a contractible subset of $S^2$ because the new $\overline{R_3}$ is
contractible in $T^2$. See figure (\ref{M_1_and_M2_m=1.99}) for an example with $m=1.99$. Recall that for $\vert m \vert>2$, $\mathcal{M}_{2}$ has the shape of a ring with a hole in its center (the northern pole of $S^2$) identical to the set $\mathcal{M}_{1}$. Therefore, for $\vert m \vert>2$ there is one decomposition of $\mathcal{M}$, $\overline{\mathcal{M}_{1}}\bigcup\overline{\mathcal{M}_{2}}=\mathcal{M}=S^2$, and, for $\vert m \vert<2$, there is another one, where one of the pieces ($\mathcal{M}_1$) can be continuously deformed between the two stages, but $\mathcal{M}_2$ cannot. There are 2 (related) topological changes: $\mathcal{M}$ went from being contractible, to being $S^2$ and, also, $R_3$ went from being topologically a ring to being a disk.
With this setting, we observe that the discontinuity of ${\mathcal{I}_v}_{3}$ and ${\mathcal{I}_c}_{3}$ at $m=2$ (as can be seen in figure (\ref{vol-vs-m})) is due to the changes in the topology of ${R_3}$. 

Note that in ${R}_{2a}\bigcup{R}_{3}$ as well as in ${R}_{2b}\bigcup{R}_{3}$, $\bar{f}$ reverses the orientation of $T^2$ with respect to $S^2$. This follows from the fact that ${\mathcal{I}_c}_{2a}+{\mathcal{I}_c}_{3}={\mathcal{I}_c}_{2b}+{\mathcal{I}_c}_{3}=-1$. As a final remark on this model in this regime, we observe that any one of the open sets ${R}_{2a}\bigcup{R}_{3}$ or ${R}_{2b}\bigcup{R}_{3}$ can be seen as a parametrization of $S^2$ with the exception of a single curve with measure zero, the boundary of $\overline{\mathcal{M}_{1}}$. Note that the quantum volume of these regions correspond to the minimal volume, $v_{min}=2$, see figure (\ref{v_2a3_vs_m}). Therefore, ${R}_{2a}\bigcup{R}_{3}$ and ${R}_{2b}\bigcup{R}_{3}$ are regions of minimal volume contained in the Brillouin torus. This is a close analogy with the spherical parametrization of $S^2$.

\subsection{Haldane model}\label{haldane-model}
The first tight-binding model of a topological insulator was introduced by Haldane in reference [\onlinecite{haldane}]. It describes spinless fermions on a honeycomb lattice where the time-reversal symmetry of the Hamiltonian is broken by a nonuniform magnetic flux per unit cell. 
By choosing the basis $a_1=(\sqrt{3},0)$ and $a_2=(\sqrt{3}/2,3/2)$ for the lattice vectors in units of the bond length, the reciprocal lattice vectors are  $b_1=\frac{2\pi}{3}(\sqrt{3},-1)$ and $b_2=\frac{4\pi}{3}(0,1)$. Let $k'=(k_x,k_y)$ be the coordinates describing the rhomboidal Brillouin torus generated by $b_1$ and $b_2$. The model is given by 
\begin{eqnarray}\label{haldane-rombic-model}
 h^{0}(k') &=&  \beta[ \mbox{cos}(k'a_1)+\mbox{cos}(k'a_2)+ \mbox{cos}(k'(a_1-a_2))]\nonumber\\
 h^{1}(k') &=& t_1 [ 1 + \mbox{cos}(k'a_1)+\mbox{cos}(k'a_2)] \nonumber\\
 h^{2}(k') &=& t_1 [  \mbox{sin}(k'a_1)+\mbox{sin}(k'a_2)]\\
 h^{3}(k') &=& m + \lambda[ \mbox{sin}(k'a_1)-\mbox{sin}(k'a_2)- \mbox{sin}(k'(a_1-a_2))]\nonumber,
\end{eqnarray}
where $t_1$ is the nearest-neighbor hopping parameter. $\beta=2t_2\mbox{cos}(\phi)$ and $\lambda=2t_2\mbox{sin}(\phi)$ include the next-nearest-neighbor hopping parameter $t_2$ where $\phi$ is the phase $t_2$ acquires, $t_2\rightarrow e^{\pm i\phi}t_2$, due to the nonuniform magnetic field. 

\begin{figure}[tbp]
\includegraphics[scale=.50]{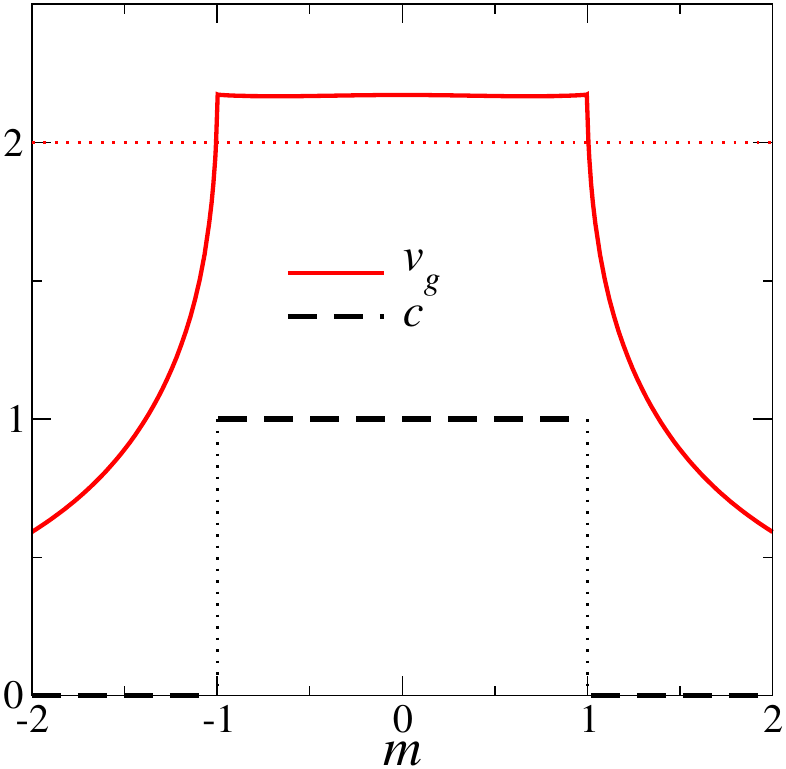}
\caption{Color online. Chern number, $c$, and quantum volume, $v_g$, as functions of $m$. Topological trivial (non-trivial) phase corresponds to $\vert m \vert >1$ ($\vert m \vert<1$). Other parameters are fixed to $t_1=1$, $t_2=\frac{1}{3\sqrt{3}}$ and $\phi=\pi/2$.}
\label{deg_vol_haldane}
\end{figure}

The change of coordinates $k^1 = \sqrt{3}k_x-\pi$, $k^2 = \frac{\sqrt{3}}{2}k_x+\frac{3}{2}k_y-\pi$ allows us to write the model over $T^2$ as follows
\begin{eqnarray}\label{haldane-model}
 h^{0}(k) &=& \beta[ -\mbox{cos}(k^1)-\mbox{cos}(k^2)+ \mbox{cos}(k^1-k^2)]\nonumber\\
 h^{1}(k) &=& t_1 [ 1 - \mbox{cos}(k^1)-\mbox{cos}(k^2)] \nonumber\\
 h^{2}(k) &=& -t_1 [  \mbox{sin}(k^1)+\mbox{sin}(k^2)]\\
 h^{3}(k) &=& m - \lambda[ \mbox{sin}(k^1)-\mbox{sin}(k^2)+ \mbox{sin}(k^1-k^2)]\nonumber.
\end{eqnarray}
Note that $h^{i}(k')=h^{i}(k'+G)$ for $G=n_1 b_1 + n_2 b_2 $ with $n_i\in\ZZ$.
As a function of $m$ and $\lambda$ the model exhibits the well known phase diagram described in the original work of Haldane. \cite{haldane}. For the purpose at hands we fix $t_1=1$, $t_2=\frac{1}{3\sqrt{3}}$. Accordingly, figure (\ref{deg_vol_haldane}) shows the Chern number and the quantum volume as functions of  $m$. The system is in a non-trivial (trivial) phase for $\vert m \vert < 1$ ($\vert m \vert > 1$).

\begin{figure}[tbp]
\centering
   \scalebox{1.0}{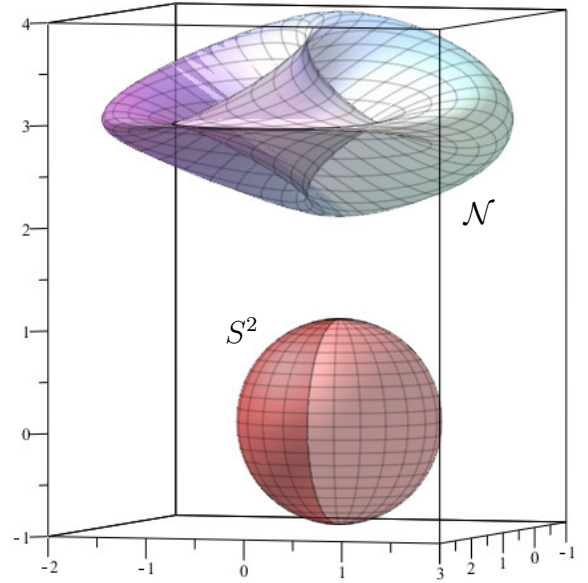}
   \caption{Color online. Subset $\mathcal{N}$ for the Haldane model for $m=3$ and $S^2$. }
\label{n-S2-for-m-3}
\end{figure}

\begin{figure}[tbp]
\includegraphics[scale=.50]{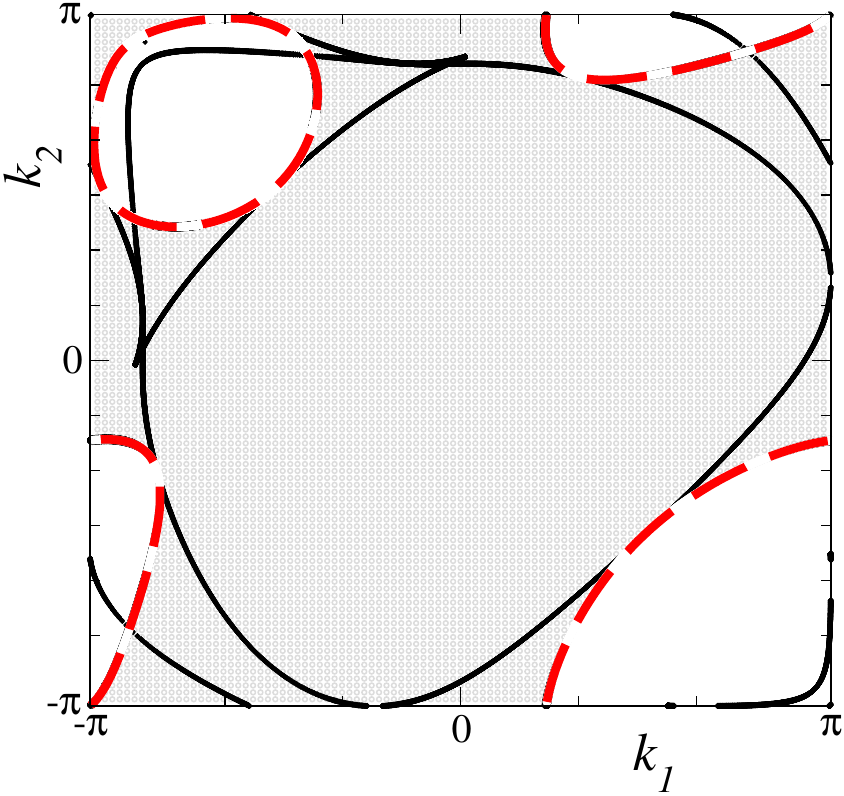}
\includegraphics[scale=.50]{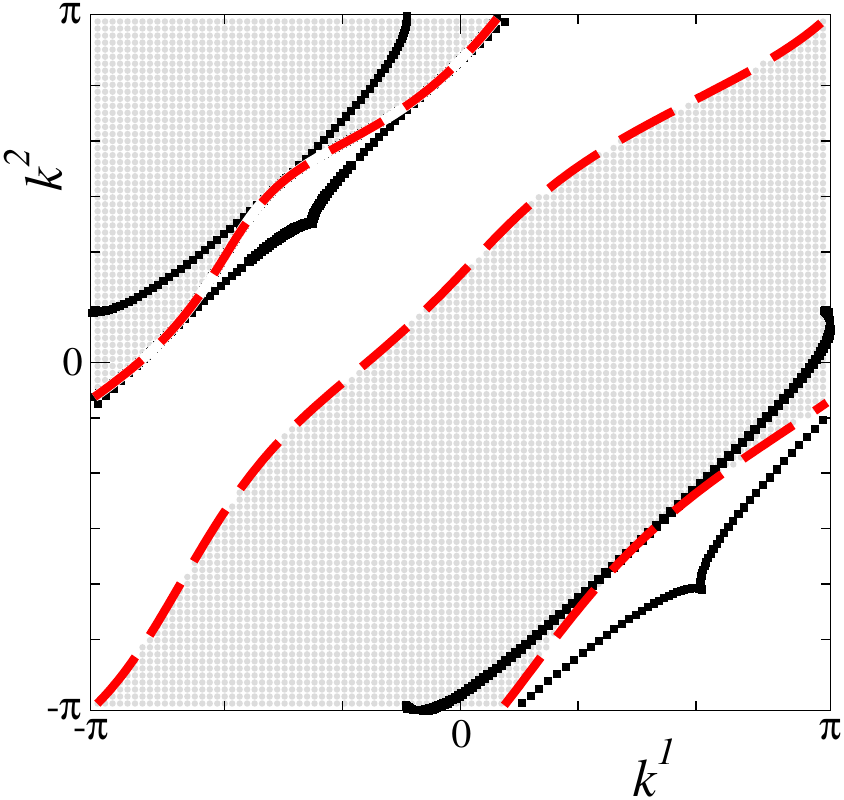}
\caption{Color online. Red dashed lines stand for $d_{\mathcal{N}}=0$. Black dotted lines are the additional curves coming from Eq. (\ref{lifting}). Grey (white) area is the region in which $\Omega_{12}(k)>0$ ($\Omega_{12}(k)<0$). Top and bottom panels are for $m=-0.5$ and $m=-2$ respectively and $\phi=\pi/2$.}
\label{haldane_regions}
\end{figure}

In figure (\ref{n-S2-for-m-3}) the set $\mathcal{N}$ for $m=3$ is shown together with $S^2$. 
We show in figure (\ref{haldane_regions}) the result of $d_{\mathcal{N}}=0$ as red dashed lines for $m=-0.5$ and $m=-2$ in the top and bottom panels respectively. The additional curves coming from Eq. (\ref{lifting}) are shown as black lines. Furthermore, for later purposes we also show the regions where $\Omega_{12}(k)>0$ ($\Omega_{12}(k)<0$) as gray (white) areas. As it can be seen from these two values of $m$, the number of curves as well as of regions in $T^2$ generated by the splitting procedure increases significantly in comparison with the previous two models we have discussed. While the integrals ${\mathcal{I}_c}_{s}$ and ${\mathcal{I}_v}_{s}$ can be straightforwardly obtained, here we simplify the presentation and give a broad discussion of the underlying geometry of the model.

\begin{figure}[tbp]
\includegraphics[scale=.30]{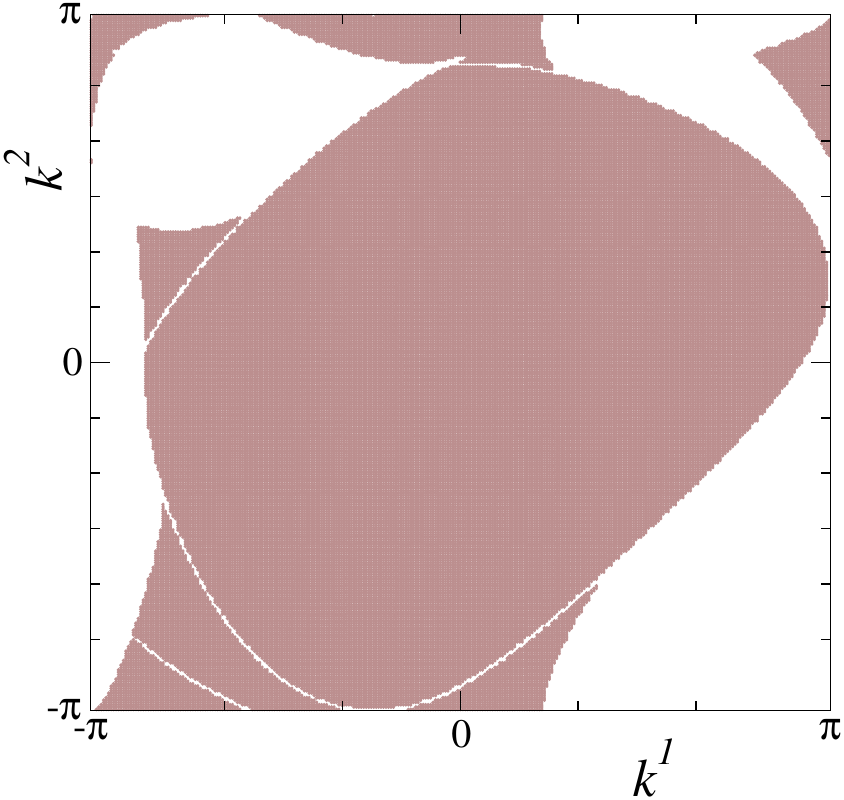}
   \scalebox{0.60}{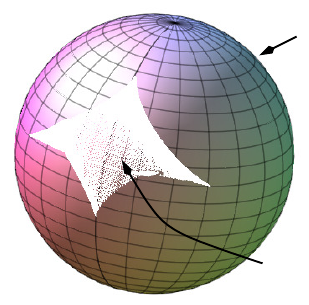}
\caption{Color online. Left: Brown area: A possible choice for ${U}_{1}$ where $\bar{f}$ is orientation-preserving ($\Omega_{12}(k)>0$) and $\bar{f}(\overline{U_{1}})=S^2$. Right: In white $\bar{f}(U_2)=\bar{f}(U_3)=\mathcal{M}_{2}\subset S^2$ and its complement $\bar{f}(R_1)=\mathcal{M}_{1}$. $m=-0.5$ and $\phi=\pi/2$. }
\label{region-topo-haldane}
\end{figure}

Notice that $v_g>2$ in the topological phase, see figure (\ref{deg_vol_haldane}). In this phase we have $\bar{f}(T^2)=\mathcal{M}=S^2$. Therefore, there must be sectors in $T^2$ that are mapped onto the same portion of $S^2$. In fact, there is an open region of minimal volume $U_{1}\subsetneq T^2$ (a single or a collection of some $R_s$) where $\bar{f}$ is injective and whose image is $S^2$ (minus a zero measure set). As a consequence, the excess of volume $v_{ex}=v_g-2$ corresponds to the volume of $T^2\setminus U_{1}$ made up by other regions in $T^2$ with the same image in $S^2$ and where $\bar{f}$ either preserves or reverses the orientation in such a way that they have no influence on the value of $c=1$. Note that at least two of those regions must be present. As in the previous model, there are different possibilities for the choice of such regions. One possibility for $U_{1}$ where $\bar{f}$ is orientation-preserving and $\bar{f}(\overline{U_{1}})=S^2$ is shown in the left panel of figure (\ref{region-topo-haldane}). By calculation, we have verified that ${\mathcal{I}_c}(U_{1})=1$ and ${\mathcal{I}_v}(U_{1})=2$. Note that $U_{1}$ is defined by curves coming from Eq. (\ref{lifting}) showing that a simple decomposition based on sign changes of $\Omega_{12}(k)$ is not sufficient for isolating a region of minimal volume in $T^2$.

On the other hand, let $U_2$ be the region made up by the gray areas of the top panel of figure (\ref{haldane_regions}) which are not included in $U_1$. Similarly, let $U_3$ be the white areas of the same figure. The subset $\bar{f}(U_2)=\bar{f}(U_3)=\mathcal{M}_{2}$ is shown in the right panel of figure (\ref{region-topo-haldane}). Note that $\bar{f}$ on $U_2$ and on $U_3$ is not injective. Furthermore, let $R_1$ be the connected region containing the origin of $T^2$ in the top panel of figure (\ref{haldane_regions}) and $\bar{f}(R_1)=\mathcal{M}_{1}$. We obtain $\mathcal{M}_{1}=S^2\setminus\overline{\mathcal{M}_{2}}$ and that $\bar{f}$ is injective on $R_1$, see right panel of figure (\ref{region-topo-haldane}). By numerical calculation, ${\mathcal{I}_c}(U_{2})=-{\mathcal{I}_c}(U_{3})\approx0.044$ while ${\mathcal{I}_v}(U_{2})={\mathcal{I}_v}(U_{3})\approx0.088$ being the excess $v_{ex}\approx0.176$. As expected, all regions contribute to the total value  $v_g\approx2.176$. 

\begin{figure}[tbp]
\centering
   \scalebox{1.0}{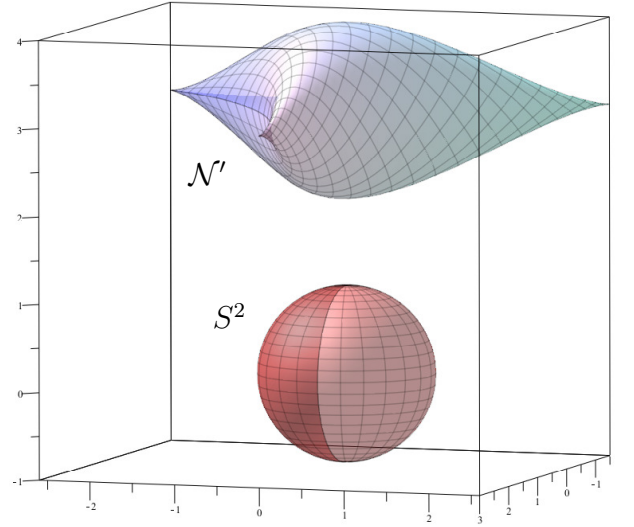}
\caption{Color online. Subset $\mathcal{N'}$ for the Haldane model with $H'_{12}(k')=h'^{1}(k')-ih'^{2}(k')=t_{1}\sum_{j=1}^{3}\mbox{exp}(-ik'e_{j})$ for $m=3$ and $S^2$.}
\label{n-S2-for-m-3-mera}
\end{figure}

We end this section with a brief comment about the two usual ways to describe Hamiltonians on the honeycomb lattice. As explained in references [\onlinecite{fruchart}] and [\onlinecite{carpentier}], there are different conventions for Bloch Hamiltonians on non-Bravais lattices where the associated Berry curvatures differs. Up to now we have used a periodic Bloch matrix, $H(k')=H(k'+G)$ for $k'\in\RR^2$ rather than on the torus, which is obtained when the trivialization of the Bloch bundle is given by Fourier transform of functions localized at the lattice points. On the other hand, if the trivialization of the Bloch bundle is performed by using periodic functions in the crystal, the Bloch Hamiltonian, $H'(k')$, may be non-periodic, $H'(k')\ne H'(k'+G)$. This description of the system appears when the off diagonal component of $H'(k')$ is written as $H'_{12}(k')=h'^{1}(k')-ih'^{2}(k')=t_{1}\sum_{j=1}^{3}\mbox{exp}(-ik'e_{j})$ where $e_{1}=(0,1)$, $e_{2}=(-\sqrt{3}/2,-1/2)$ and $e_{3}=(\sqrt{3}/2,-1/2)$ are the vectors connecting a given site to its three nearest neighbors. The diagonal components do not change, so $h'^{0}(k')=h^0(k')$ and $h'^{3}(k')=h^3(k')$. In this case the set $\mathcal{N'}=F'_{1}(T^2)$, with $F'_{1}(k)=(h'^1(k),h'^2(k),h'^3(k))$, becomes the surface shown in figure (\ref{n-S2-for-m-3-mera}).

\begin{figure}[tbp]
\includegraphics[scale=.63]{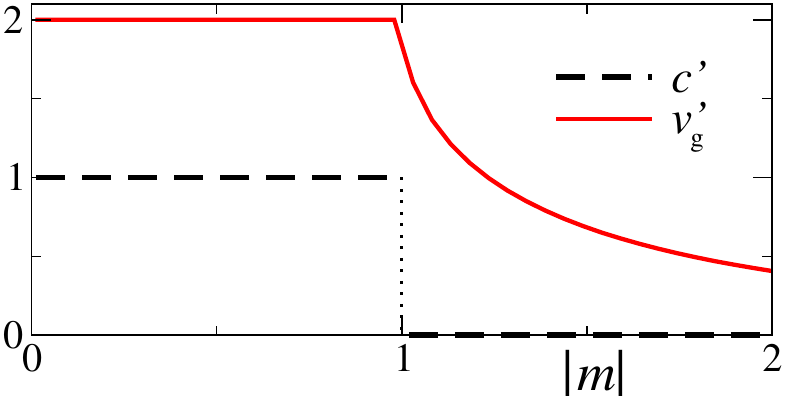}
\includegraphics[scale=.30]{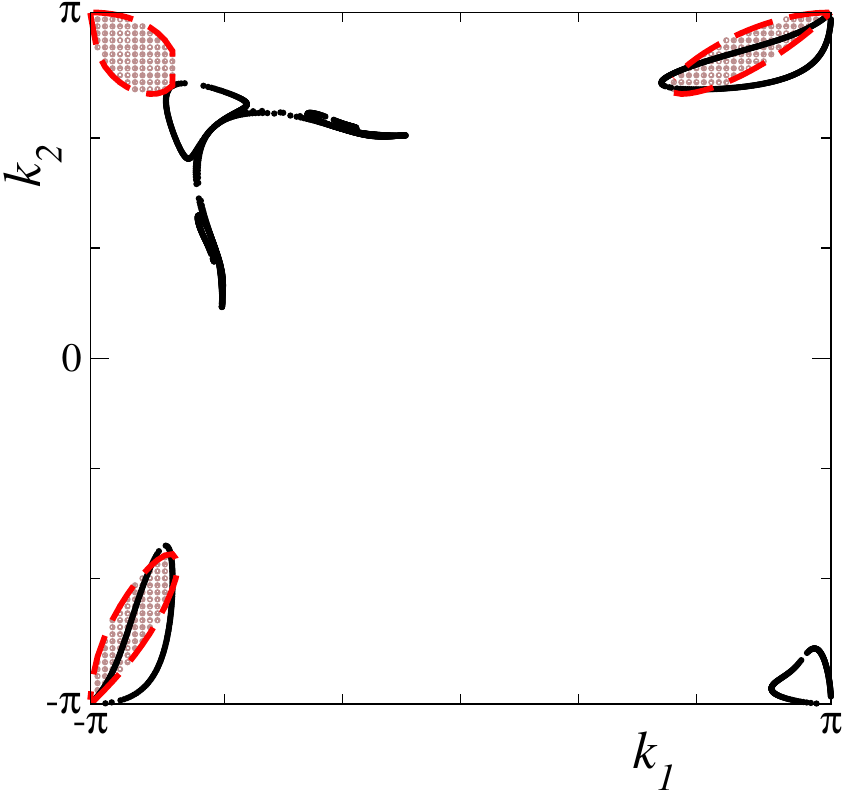}
\includegraphics[scale=.30]{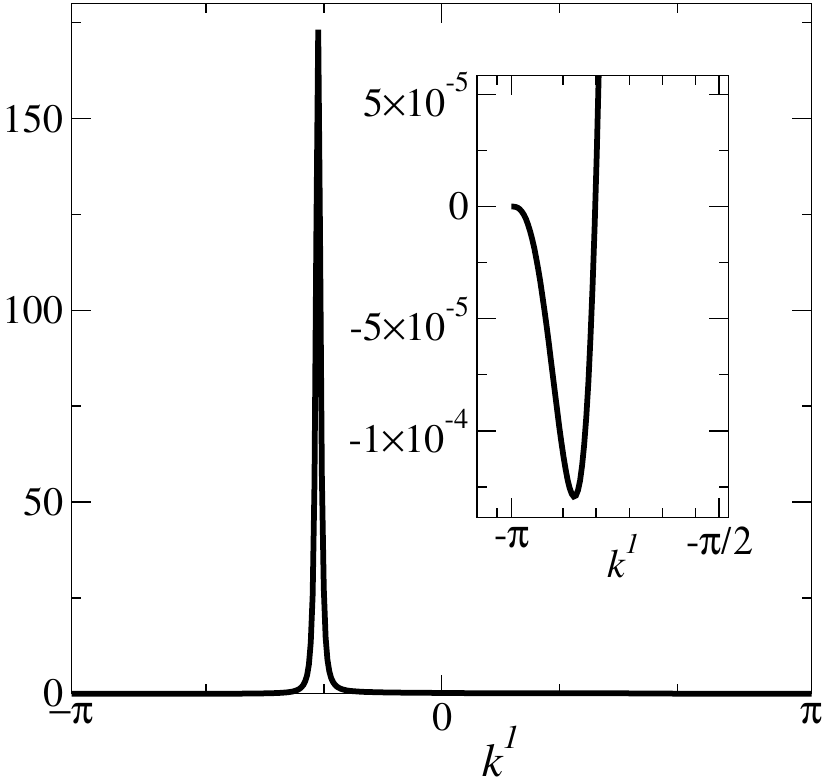}
\caption{Color online. Top panel: Chern number, $c'$, and quantum volume, $v'_g$, as functions of $m$ for the Haldane model with $H'_{12}(k')=h'_{1}(k')-ih'_{2}(k')$. Other parameters are fixed to $t_1=1$, $t_2=\frac{1}{3\sqrt{3}}$ and $\phi=\pi/2$. Bottom left panel: Red dashed lines stand for $d_{\mathcal{N'}}=0$ for $m=-0.99$. Black dotted lines are the additional curves $ \gamma'$ coming from Eq. (\ref{lifting}). Brown (white) area is the region where $\Omega'_{12}(k)<0$ ($\Omega'_{12}(k)>0$). Bottom right panel: $\Omega'_{12}(k)$ as a function of $k^1$ with $k^2=-k^1$. The inset shows a region where $\Omega'_{12}(k)<0$.}
\label{berry-mera}
\end{figure}

Let $d_{\mathcal{N'}}(k)$ be the function analogous to the one defined in Eq. (\ref{det-N}) but written in terms of $h'^{i}(k)$. Furthermore, let $\Omega'_{12}(k)$ be the associated Berry curvature and $c'$ the corresponding first Chern number. As $\Omega_{12}(k)$ and $\Omega'_{12}(k)$ are the curvature of two connections on the same bundle -the Bloch bundle- their integrals over $T^2$ compute the Chern number of the bundle so that $c'=c$. We have verified that $c'=c$ as functions of $m$, see figure (\ref{berry-mera}), APPENDIX \ref{berry-conventions} and references [\onlinecite{fruchart}] and [\onlinecite{carpentier}]. The same figure also shows the values of the quantum volume $v'_g$ as a function of $m$. The first thing to note is that, comparing with figure (\ref{deg_vol_haldane}), $v'_g(m)\ne v_g(m)$. This is also an expected result. Since $\mathcal{N'}\ne\mathcal{N}$ then $\mathcal{M'}$ and $\mathcal{M}$ do not have to agree in general. In particular their volumes do not have to agree and this is indeed the case. Second, from figure (\ref{berry-mera}) it seems, at first sight, that inside the topological phase $v'_g=2c'$. If this were the case, $v'_g$ acquires a topological character and also $d_{\mathcal{N'}}(k)\ge0$ for all $k\in T^2$. This is to say that the Berry curvature, $\Omega'(k)$, does not change sign inside the Brillouin torus pointing to a single region (up to a zero measure set, $d_{\mathcal{N'}}(k)=\Omega'(k)=0$) in the torus. However, as it is well known, both of these statements are false.  

Focusing on the topological phase, the results of $d_{\mathcal{N'}}(k)=0$ and Eq. (\ref{lifting}) are shown in the bottom left panel of figure (\ref{berry-mera}) for $m=-0.99$. The brown area in the figure is the region where $\Omega'_{12}(k)<0$ while the white one corresponds to $\Omega'_{12}(k)>0$. Accordingly, the same analysis that we have done in the periodic convention using some sets $U'_1$, $U'_2$ and $U'_3$ holds. In particular, a non vanishing $v'_{ex}$ is expected. Remarkably, while $\Omega'_{12}(k)<0$ in an macroscopic area of $T^2$, its intensity over these regions is very low compared to the maximum of $|\Omega'_{12}(k)|$. In the bottom right panel of figure (\ref{berry-mera}) we show $\Omega'_{12}(k)$ as a function of $k^1$ along the line $k^2=-k^1$. A closer inspection of $\Omega'_{12}(k)$ shows the changing of sign, see the inset of the figure. Within the accuracy of numerical error we obtain $c'\approx0.99999999$, $v'_g\approx2.00004314$ and then $v'_{ex}\approx0.00004314$. We stress that no matter how small the value of $v'_{ex}$ is, from the topological point of view $v'_g$ is, as expected, not endowed with a topological meaning and $v'_g\ne2c'$. 
Similar behavior is present for $m$ in the range $0<|m|<1$ and other model parameters. 
Finally, it should be noted that [\onlinecite{ozawa}] states incorrectly that $v'_g=2c'$ probably because of the small negative value of $\Omega'_{12}(k)$, for the presented model parameters.

\section{SUMMARY AND Conclusions}\label{sum}
We have studied some geometric properties of two-dimensional two-band topological insulators that complement the topological ones. Within the topologically non-trivial phase of a given model describing these insulators, the value of the Chern number $ c = n$, with $n$ an integer, asserts that $S^2$ is wrapped around, at least, $n$ times by the classifying map $\bar{f}$ taking orientations into account. However, it tells us nothing about the presence of extra, full or partial, wrappings with opposite orientations canceling each other out and, therefore, preserving the value of $c$. This information is contained in the quantum volume $v_g$. Such extra wrappings, if they are present, invalidate a topological interpretation for $v_g$ leading to the inequality $v_g>2c$.

In this work we have introduced a systematic procedure of extracting this information from $v_g$. We have presented a procedure that splits the Brillouin torus in connected open regions, $R_s$, such that $\bar{f}$ restricted to each one of them is a local diffeomorphism with an open subset of $S^2$. A direct consequence is that in each one of these regions the quantum metric is positive definite. Moreover, within each open region $R_s$, there is a natural relation between the integrals contributing to $v_g$ and $c$, ${\mathcal{I}_v}_{s}=\pm2{\mathcal{I}_c}_{s}$ where the sign is that of the Berry curvature in the region and is a consequence of the preserving/reversing orientation character of $\bar{f}$. The regions $R_s$ are bounded by the curves along which the Berry curvature vanishes together with the ones that are mapped to the same set of points within the two-sphere by the classifying map.  

In particular, the procedure is able to isolate regions of minimal volume in the Brillouin zone. The integrals of the square root of the determinant of the quantum metric and the Berry curvature over that region give $v_{min}$ and $c$ respectively.

We have illustrated these ideas by using a time-reversal two-dimensional two-band system, and two well known models for Chern insulators, the simplest on a square lattice and the Haldane model. The three examples are qualitatively similar under the analysis. For each of them, not only the values of ${\mathcal{I}_v}_{s}$ and ${\mathcal{I}_c}_{s}$ are calculated but also complete details of the underlying geometric structures from which they emerge are given. We provided graphs of each region $R_s$ in the Brillouin torus as well as of its image under the classifying map in $S^2$ and $\RR^3$, including its shape, volume and the sign of the Berry curvature.  

Although the examples that we have considered have an absolute value of the Chern number equal to one, in the topological phase, the procedure can be applied to more general Chern insulators.

Finally, we have discussed and clarified an incorrect interpretation of $v_g$ as an Euler characteristic number that appears in the literature.

\textit{Acknowledgment}. 
We are partially supported by CONICET, Argentina. 
P. R. B, was sponsored by PICT 2013-1045 of the ANPCyT and PIP 364 of CONICET.
J. F. and F. K. were partially supported by the Universidad Nacional de Cuyo through grants 06/C009-T1 and 06/C567. 

\appendix
\section{Expression for the classifying map}\label{classifying_f}
In this appendix we prove the expression that appears in Eq. (\ref{eq:f}) for the classifying map $\bar{f}$.
\subsubsection{The Bloch Valence Bundle}\label{sec:the_physical_bundle}

As in the main text  we consider the trivial
$\CC$-vector bundle of rank $2$ over $T^2$ ($p_1:T^2\times \CC^2\rightarrow T^2$),
with its canonical hermitian metric. From Eq. (\ref{chern-ham}) with $h^0(k)=0$, the Bloch Hamiltonian $H(k)$ can be seen as a hermitian endomorphism of this bundle. For each $k\in T^2$, the eigenvalues of $H(k)$ are
\begin{equation*}
  \lambda_\pm(k)= \pm \norm{h(k)}
  =\pm\sqrt{(h^1(k))^2+(h^2(k))^2+(h^3(k))^2}.
\end{equation*}
We assume, in what follows, that $\norm{h(h)}\neq 0$, so that the two
eigenvalues of $H(k)$ are different. The corresponding eigenspace for the lowest eigenvalue is
\begin{equation*}
  {\mathcal V}_k =
  \begin{cases}
    \left< \left(
      \begin{array}{c}
        h^1(k)-i h^2(k),\\
        \lambda_{-}(k)-h^3(k)
      \end{array}
    \right) \right>,\quad \text{ if } h^3(k) \neq \lambda_{-}(k),\\
    \left< \left(
      \begin{array}{c}
        h^3(k)+\lambda_{-}(k),\\
        h^1(k)+i h^2(k),
      \end{array}
    \right) \right>,\quad \text{ if } h^3(k) \neq -\lambda_{-}(k)
  \end{cases}
\end{equation*}
Notice that, as $\lambda_{-}(k)\neq 0$, the two cases cover all of
$T^2$. Furthermore, it is easy to check that the two descriptions
match on the overlapping region, so that the definition is
consistent. Hence $ {\mathcal V}\rightarrow T^2$ is a
$\CC$-vector bundle of rank $1$. Thus we have
\begin{equation*}
  \tilde{f}:T^2\rightarrow \CC\PP^1 \stext{ defined by }
  \tilde{f}(k) = {\mathcal V}_k\subset \CC^2.
\end{equation*}
It follows that ${\mathcal V} \simeq \tilde{f}^*\Theta$ with $\Theta$ the tautological line bundle over ${\CC\PP^1}$. Schematically:
\begin{equation*}
  \xymatrix{ {{\mathcal V}^\pm} \ar[dr] \ar[r]^{\simeq} &
    {\tilde{f}^*\Theta} \ar[d]^{p_1} \ar[r]^{p_2}& {\Theta} \ar[d]\\
    {} & {T^2} \ar[r]_{\tilde{f}_\pm} & {\CC\PP^1}
  }
\end{equation*}


\subsubsection{$\CC\PP^1$ and $S^2$}
\label{sec:CP1_and_S2}

There is a well known identification of $\CC\PP^1$ and
$\overline{\CC} = \CC\cup \{\infty\}$ (the extended complex plane). The mapping is
\begin{equation*}
  \eta:\CC\PP^1\rightarrow \overline{\CC} \stext{ for }
  \eta(\langle \svec{z^0}{z^1}\rangle) =
  \begin{cases}
    z^0/z^1 \text{ if } z^1\neq 0,\\
    \infty, \text{ if } z^1=0.
  \end{cases}
\end{equation*}

Last, consider $S^2=\{u\in\RR^3:\norm{u}=1\}$. Fixing a point
$p\in S^2$, we have the stereographic projection from $p$:
$St_p:S^2\SM\{p\}\rightarrow \CC$ so that $St_p(u)$ is the point of
intersection between the real line that runs through $p$ and $u$ with
the plane containing $0\in \RR^3$ perpendicular to $\langle p \rangle$, that we identify with $\CC$. 
It is easy to check
that $St_p$ is a diffeomorphism; furthermore, by choosing
two different points of projection it happens that the two maps define
compatible charts identifying $S^2$ with $\overline{\CC}$. In
particular, when $p=(0,0,1)$, the North Pole of $S^2$, we have
\begin{equation*}
  St_N(u^1,u^2,u^3) = \frac{u^1+i u^2}{1-u^3}.
\end{equation*}

Putting these two constructions together with the classifying map one has the following diagram

\begin{equation*}
  \xymatrix{
    {} & {\Theta}\ar[d]\\
    {} & {\CC\PP^1} \ar[d]_{\eta} \\
    {S^2} \ar[r]^{St} & \overline{\CC} \\
    {S^2\SM\{N\}} \ar@{^{(}->}[u]^{i_{S^2}} \ar[r]_-{St_N}
      & {\CC} \ar@{^{(}->}[u]^{i_{\overline{\CC}}}
  }
\end{equation*}
All together, we can show a big diagram together with the Bloch valence bundle:
\begin{equation*}
  \xymatrix{
    {{\mathcal V}} \ar[d] & {\tilde{f}^*\Theta} \ar[l]_{\simeq} \ar[dl]
    &{} & {\Theta}\ar[d]\\
    {T^2} \ar@/_1.5pc/[rrr]^(.4){\tilde{f}} \ar@{-->}@/_/[rrd]_{f}
    & {} &{} &  {\CC\PP^1} \ar[d]_{\eta} \\
    {} & {} & {S^2} \ar[r]_{St} & \overline{\CC}
  }
\end{equation*}
where $f = St^{-1}\circ \eta\circ \tilde{f}$.  Thus, if
$h^3(k)\neq \lambda_{-}(k)$, we have
\begin{equation*}
  \begin{split}
    {f}(k) =& St^{-1}(\eta(\tilde{f}(k))) \\=&
    St^{-1}\left(\eta\left(\left<\left(
          \begin{array}{c}
            h^1(k)-ih^2(k)\\\lambda_{-}(k)-h^3(k)
          \end{array}
        \right)\right>\right)\right) \\=& St^{-1}
  \left(\frac{h^1(k)-ih^2(k)}{\lambda_{-}(k)-h^3(k)}\right) = St^{-1}
  \left(\frac{\frac{h^1(k)}{\lambda_{-}(k)}+i\frac{-h^2(k)}{\lambda_{-}(k)}}{1-\frac{h^3(k)}{\lambda_{-}(k)}}\right)
  \\=&
  St^{-1}\left(St\left(\frac{(h^1(k),-h^2(k),h^3(k))}{\lambda_{-}(k)}\right)\right) \\=&
  \frac{(h^1(k),-h^2(k),h^3(k))}{\lambda_{-}(k)}
  \end{split}
\end{equation*}

If we define
\begin{gather*}
  F_1:T^2\rightarrow X \stext{ by } F_1(k)=(h^1(k),h^2(k),h^3(k)),\\
  F_2:X\rightarrow S^2 \stext{ by } 
  F_2(x^1,x^2,x^3)=\frac{(-x^1,x^2,-x^3)}{\norm{(-x^1,x^2,-x^3)}},
\end{gather*}
we see that ${f}(k) = (F_2\circ F_1)(k)$. 

\section{Pullbacks of the Berry curvature and metric tensor from $S^2$ to the Brillouin torus}\label{pullback}

\textit{Berry curvature.}
Let $S^2\subset \RR^3$ oriented by the exterior normal. The coordinate chart given by the spherical coordinates $(\theta,\phi)$, with $0<\theta<\pi$ and $0<\phi<2\pi$, is positively oriented with respect to this orientation. Accordingly, $\omega_{S^2}=d\theta\wedge d\phi$ is an orientation form for the chosen orientation. The Berry curvature on the tautological line bundle $\Theta$ over $\CC\PP^1\simeq S^2$ is a two-form $\tilde\Omega$ on the base and in the spherical coordinate chart it is given by ~\cite{bernevig-book}
\begin{eqnarray}\label{berry-curvature-over-S2}
 \tilde{\Omega}=\frac{1}{2}\mbox{sin}\theta~ d\theta\wedge d\phi
\end{eqnarray}
where the factor $1/2$ comes from the isometry between $\CC\PP^1$ with its Fubini-Study metric and the sphere of radius $1/2$ in $\RR^3$ with its round metric.\cite{madsen-book} In what follows we will denote the sphere of radius $1/2$ by $S^2$. Therefore 
$\tilde{\Omega}=\lambda\omega_{S^2}$, with $\lambda=\frac{1}{2}\mbox{sin}\theta>0$ in $\theta\in(0,\pi)$, provides an equivalent orientation form on $S^2$.

On the other hand, the following two-form in $\RR^3$ is nowhere zero when restricted to $S^2$,
\begin{eqnarray}\label{berry-curvature-over-S2-in-R3}
 \tilde{\Omega}'&=&\frac{1}{2}\big[ x^{1}dx^{2}\wedge dx^{3} + 
                              x^{2}dx^{3}\wedge dx^{1} +
                              x^{3}dx^{1}\wedge dx^{2} \big]\nonumber\\
                &=&\frac{1}{4}\sum_{ijk}\epsilon_{ijk}  x^{i}dx^{j}\wedge dx^{k}.           
\end{eqnarray}
Moreover, $\tilde{\Omega}=i^{\ast}_{S^2}\tilde{\Omega}'$, where $i_{S^2}$ is the inclusion of $S^2$ in $\RR^3$. 

When pulled back by $\bar{f}:T^2\rightarrow S^2$ to $T^2$, it gives rise to the following two-form 
\begin{eqnarray}\label{berry-curvature-over-T2}
 {\Omega'}&=&\bar{f}^{\ast}\tilde{\Omega}'=\Omega'_{12}(k)~dk^{1}\wedge dk^{2},            
\end{eqnarray}
where $\Omega'_{12}(k)$ is  
\begin{eqnarray}\label{componente-berry-curvature-over-T2}
 \Omega'_{12}(k)&=&\frac{1}{4}\sum_{ijk}\epsilon_{ijk}~x^{i}(k)\sum_{\mu\nu}  
               \epsilon^{\mu\nu}
               \frac{\partial x^{j}(k)}{\partial k^{\mu}} 
               \frac{\partial x^{k}(k)}{\partial k^{\nu}}   \nonumber\\
               &=&\frac{1}{4}\sum_{ijk}\epsilon_{ijk}~x^{i}(k)  
               \Big[\frac{\partial x^{j}(k)}{\partial k^{1}} 
                \frac{\partial x^{k}(k)}{\partial k^{2}}-\frac{\partial x^{k}(k)}{\partial k^{1}}\frac{\partial x^{j}(k)}{\partial k^{2}} 
                \Big]   \nonumber\\
                &=&\frac{1}{2}\sum_{ijk}\epsilon_{ijk}~x^{i}(k)  
               \frac{\partial x^{j}(k)}{\partial k^{1}} 
                \frac{\partial x^{k}(k)}{\partial k^{2}},
\end{eqnarray}
where $x^{i}(k)=\bar{F}^{i}(k)=h^{i}(k)/\norm{h(k)}$. Finally,
\begin{eqnarray}\label{componente-berry-curvature-over-T2-2}
 \Omega'_{12}(k)&=&\frac{1}{2\norm{h(k)}^{3}}\sum_{ijk}\epsilon_{ijk}~
                 h^{i}(k)\frac{\partial h^{j}(k)}{\partial k^{1}}
                          \frac{\partial h^{k}(k)}{\partial k^{2}}.~~~~~~~~
\end{eqnarray}
Note that $\Omega'_{12}(k)=\Omega_{12}(k)$ (see Eq. (\ref{berry-curvature})) and then $\Omega'=\Omega$, so the Berry curvature defined from the quantum geometric tensor as $2\mbox{Im}Q$ is nothing but the pullback of $\tilde{\Omega}$ to $T^2$.

\vspace{0.5cm}
\textit{Riemannian metric.}
The metric tensor $\tilde{g}$ for the sphere $S^2$ of radius $1/2$ embedded in $\RR^3$ in the same spherical coordinate chart is given by 
\begin{eqnarray}\label{metric-in-S2-in-R3}
 \tilde{g}&=&\frac{1}{4}d\theta\otimes d\theta + 
             \frac{1}{4}\mbox{sin}^{2}(\theta)d\phi\otimes d\phi.
\end{eqnarray}
Similar to the Berry curvature, $\tilde{g}$ can be seen as the restriction to $S^2$ of the following tensor in the ambient space $\RR^3$
\begin{eqnarray}
 \tilde{g}'&=&\sum_{i}\frac{1-(x^{i})^{2}}{4}dx^{i}\otimes dx^{i} - 
             \sum_{i<j}\frac{x^{i}x^{j}}{2} dx^{i}\otimes dx^{j}.~~~
\end{eqnarray}
When it is pulled back to the Brillouin torus by $\bar{f}$, it gives rise to the symmetric tensor 
\begin{eqnarray}\label{metric-in-T2}
g'&=&\sum_{\mu\nu}g'_{\mu\nu}(k)~dk^{\mu}\otimes dk^{\nu},\nonumber
\end{eqnarray}
where the components as functions of the momentum coordinates in the torus  are given by, 
\begin{eqnarray}\label{componentes}
g'_{\mu\nu}(k)&=&\sum_{i}a^{i}(k)~\frac{\partial x^{i}(k)}{\partial k^{\mu}}~\frac{\partial x^{i}(k)}{\partial k^{\nu}}\\
&-&\sum_{i<j}b^{ij}(k)\big( 
\frac{\partial x^{i}(k)}{\partial k^{\mu}}~\frac{\partial x^{j}(k)}{\partial k^{\nu}}+
\frac{\partial x^{i}(k)}{\partial k^{\nu}}~\frac{\partial x^{j}(k)}{\partial k^{\mu}}
\big)\nonumber
\end{eqnarray}
with $a^{i}(k)=\frac{1-(x^{i}(k))^{2}}{4}$ and $b^{ij}(k)=\frac{x^{i}(k)x^{j}(k)}{4}$.

Similar to the Berry curvature, it can be seen that $g=g'$, that is, $g=\mbox{Re}Q$  is the pullback of the Fubini-Study metric of $\CC\PP^1$ to $T^2$. 
The functions $\Omega_{12}(k)$ and the value of $\sqrt{\mbox{det}(g(k))}$,
with $\mbox{det}(g(k))=g'_{11}(k)g'_{22}(k)-g'_{12}(k)g'_{21}(k)$ are the integrands that appear when computing $c$ and $v_g$ and their different contributions.


\section{Two different conventions for Bloch Hamiltonians}\label{berry-conventions}
In this appendix we show that the change in the Berry curvature for the Haldane model when it is written in two different conventions describing the same honeycomb lattice is given by a smooth function whose integral over $T^2$ vanishes. Furthermore, since the Berry curvature is gauge independent we can work with the element $|u'_{-}(k')\rangle= {\mbox{N}'}^{-1}(k') \left| \begin{array}{c} h'^1(k')-i h'^2(k')\\ -h(k')-h'^3(k') \end{array}\right>$ of the eigenspace ${\mathcal V}_{k'}$ where $k'=(k_x, k_y)$ are coordinates for the rhombohedrical Brillouin torus and ${\mbox{N}'}(k')=[2h(k')(h(k')+h'^3(k'))]^{1/2}$ is a normalization factor.

The non-periodical, $H'(k')$, and the periodical, $H(k')$, descriptions of the model are related by $H'_{12}(k')=\mbox{exp}(-ik'e_2)H_{12}(k')$. Therefore, in the coordinates $k=(k^1, k^2)\in[-\pi,\pi]^2$, we obtain $|u'_{-}(k)\rangle=\hat{U}(k)|u_{-}(k)\rangle$, where $\hat{U}(k)=\left(
    \begin{array}{cc}
      \mbox{exp}(-i\alpha(k))&0\\
      0&1
    \end{array}
  \right)$ is a unitary matrix and $\alpha(k)=(k^1+k^2+2\pi)/3$ and $|u_{-}(k)\rangle$ was defined in Appendix \ref{classifying_f}. As a consequence,
\begin{eqnarray}\label{partial_u_prime}
\partial_{j}|u'_{-}(k)\rangle&\equiv&|\partial_j u'_{-}(k)\rangle\\
&=&\big(\partial_{j}\hat{U}(k)\big)|u_{-}(k)\rangle+\hat{U}(k)|\partial_j u_{-}(k)\rangle\nonumber
\end{eqnarray}

From the definition in Eq. (\ref{qgt}) the Berry curvature is given by  
\begin{eqnarray}
 \Omega'_{12}(k)&\equiv&-2\mbox{Im}\big[\langle\partial_1 u'_{-}(k)|\big(\hat{1}-|u'_{-}(k)\rangle\langle u'_{-}(k)|\big)|\partial_2 u'_{-}(k)\rangle\big]\nonumber\\
 &=&-2\mbox{Im}[\langle\partial_1 u'_{-}(k)|\partial_2 u'_{-}(k)\rangle]\\
 &=&i\big( \langle\partial_1 u'_{-}(k)|\partial_2 u'_{-}(k)\rangle - \langle\partial_2 u'_{-}(k)|\partial_1 u'_{-}(k)\rangle\big)\nonumber,
\end{eqnarray}
where the first step after the definition is a consequence of $\mbox{Re}\langle\partial_j u'_{-}(k)|u'_{-}(k)\rangle=\mbox{Re}\langle u'_{-}(k)|\partial_j u'_{-}(k)\rangle=0$ and the last step follows from the fact that if $W\in\CC$ then $\mbox{Im}(W)=(W-\overline{W})/2i$.
By using Eq. (\ref{partial_u_prime}) the relation between $\Omega'_{12}(k)$ and $\Omega_{12}(k)$ can be written as 
\begin{eqnarray}
 \Omega'_{12}(k)&=&\Omega_{12}(k)+i\Big[ \langle u_{-}(k)|\hat{O}_{12}(k)|u_{-}(k)\rangle\nonumber\\
 &+&\langle u_{-}(k)|v_{12}(k)\rangle+\langle v_{12}(k)|u_{-}(k)\rangle\Big]
\end{eqnarray}
where 
\begin{eqnarray}
 \hat{O}_{12}(k)&=&\big(\partial_1\hat{U}(k)\big)^\dagger\big(\partial_2\hat{U}(k)\big)-\big(\partial_2\hat{U}(k)\big)^\dagger\big(\partial_1\hat{U}(k)\big)\nonumber\\
 |v_{12}(k)\rangle&=&\big(\partial_1\hat{U}\big)^\dagger\hat{U}|\partial_2 u_{-}(k)\rangle-\big(\partial_2\hat{U}\big)^\dagger\hat{U}|\partial_1 u_{-}(k)\rangle\nonumber.
\end{eqnarray}
The following step uses $\partial_1\alpha(k)=\partial_2\alpha(k)=1/3$ and $\langle v_{12}(k)|u_{-}(k)\rangle=\frac{i}{3\mbox{N}(k)}(h^{1}(k)+ih^{2}(k))\big(\partial_2-\partial_1\big)\big((h^{1}(k)+ih^{2}(k))/\mbox{N}(k)\big)$,
\begin{eqnarray}\label{omega_prime}
 \Omega'_{12}(k)&=&\Omega_{12}(k)+(1/6)\big(\partial_2-\partial_1\big)x^{3}(k).
\end{eqnarray}
Finally, the integral over the Brillouin torus of the last term in Eq. (\ref{omega_prime}) vanishes because $x^{3}(k)$ is a continuous function over $T^2$ (periodic in $\RR^2$). Therefore, $c'=c$. More details can be found in references [\onlinecite{fruchart}] and [\onlinecite{carpentier}].

\section{Integration over ${R}_s$}\label{integration}
In this appendix we describe a procedure for the numerical evaluation of the integrals in Eq.(\ref{cher-volume-contributions}).

Regard $T^2$ as the square $[-\pi,\pi]^2\subset\RR^2$ where the opposite edges are identified. Let $\beta$ be the set of points in $T^2$ coming from the boundary of that square. Removing $\beta$ from $T^2\setminus\gamma'$ could result in the splitting of ${R}_{s}$ into different connected components (assumed finite), $\mathcal{R}_{s,\alpha}$, that now can be seen as open sets in $\RR^2$ and in $T^2$ simultaneously. In what follows we will omit this difference. Let $i_{s,\alpha}:\mathcal{R}_{s,\alpha}\to T^2$ be the inclusion
of $\mathcal{R}_{s,\alpha}$ into $T^2$. If $\gamma\cup\beta$ is a union of finitely many curves, then it has zero measure in $T^2$. Furthermore, since $[-\pi,\pi]^2=\bigcup \overline{\mathcal{R}_{s,\alpha}}$, where $\overline{A}$ stands for the closure of the set $A$ , we obtain for a given two-form $\omega$ on $T^2$:

\begin{eqnarray}\label{integracion-en-Rs}
\int_{T^2}\omega&=&\sum_{s=1}^{N}\mathcal{I}_{s},\quad\text{where}\nonumber\\
\mathcal{I}_{s}&=&\sum_{\alpha=1}^{N_{\alpha}}\mathcal{I}_{s,\alpha}=\sum_{\alpha=1}^{N_{\alpha}}\int_{\mathcal{R}_{s,\alpha}}i^{\ast}_{s,\alpha}\omega.
\end{eqnarray}

Notice that this elementary procedure does not involve the use of partition of unity.
See also reference [\onlinecite{lee-book},\S16] for further details.


\begin{thebibliography}{99}
\bibitem{topo-review} Xiao-Liang Qi and Shou-Cheng Zhang, \textit{Topological Insulators And Superconductors}, Rev. Mod. Phys. 83 (2011) 1057-1110, arXiv:1008.2026.

\bibitem{bernevig-book} B. A. Bernevig and T. L. Hughes, \textit{Topological Insulators And Topological Superconductors},  Princeton University Press, 2013.

\bibitem{thouless} D. J. Thouless, M. Kohmoto, M. P. Nigthingale, and M. den Nijs, \textit{Quantized Hall conductance in a two-dimensional periodic potential}; Phys. Rev. Lett. \textbf{49} 405 (1982).

\bibitem{thouless-book} D. J. Thouless, \textit{Topological Quantum Numbers in Nonrelativistic Physics}, World Scientific Publishing Co. Pte Ltd. 1998.

\bibitem{fruchart-carpentier} M. Fruchart and D. Carpentier, \textit{An Introduction to topological insulators}, Comptes Rendus Physique \textbf{14} 779 (2013).

\bibitem{berry} Berry, M. V. \textit{Quantal phase factors accompanying adiabatic changes}. Proceedings of the Royal Society of London A: Mathematical, Physical and Engineering Sciences, vol. 392, 45 (1984).

\bibitem{simon} B. Simon,  \textit{Holonomy, the Quantum Adiabatic Theorem, and Berry's Phase}; Phys. Rev. Lett. \textbf{51} 2167 (1983).


\bibitem{yu-quan-ma-0}Yu-Quan Ma, Shu Chen, Heng Fan, and Wu-Ming Liu; \textit{Abelian and non-Abelian quantum geometric tensor}; Phys. Rev. B \textbf{81} 245129 (2010). 

\bibitem{provost} J. P. Provost, G. Vallee, \textit{Riemannian Structure on Manifolds of Quantum States}, Commun. Math. Phys. \textbf{76} (1980) 289.

\bibitem{wootters} W. K. Wootters, \textit{Statistical distance and Hilbert space}; Phys. Rev. D \textbf{23}, 357 (1981).

\bibitem{braunstein} S. L. Braunstein and C. M. Caves, \textit{Statistical distance and the geometry of quantum states}; Phys. Rev. Lett. \textbf{72}, 3439 (1994).

\bibitem{anandan} J. Anandan and Y. Aharonov, \textit{Geometry of quantum evolution}; Phys. Rev. Lett. 65, 1697 (1990).
\bibitem{rezakhani} A. T. Rezakhani, \textit{et al.}, \textit{Intrinsic geometry of quantum adiabatic evolution and quantum phase transitions}, Phys. Rev. A \textbf{82}, 012321 (2010).

\bibitem{sivak} D. A. Sivak and G. E. Crooks, \textit{Thermodynamic metrics and optimal paths}, Phys. Rev. Lett. \textbf{108}, 190602 (2012).

\bibitem{scandi} M. Scandi and M. Perarnau-Llobet, \textit{Thermodynamic length in open quantum systems}, Quantum \textbf{3}, 197 (2019).

\bibitem{katabarwa} A. Katabarwa \textit{et al}., \textit{Connecting geometry and performance of two-qubit parameterized quantum circuits}, Quantum \textbf{6}, 782 (2022)].

\bibitem{bhandari} B. Bhandari, \textit{et al.}, \textit{Geometric properties of adiabatic quantum thermal machines}, Phys. Rev. B \textbf{102}, 155407 (2020). 

\bibitem{tan} X. Tan, \textit{et al.}, \textit{Experimental Measurement of the Quantum Metric Tensor and Related Topological Phase Transition with a Superconducting Qubit}, Phys. Rev. Lett. \textbf{122}, 210401 (2019).

\bibitem{zhu} Yan-Qing Zhu, \textit{et al.}, \textit{Note on 'Experimental Measurement of Quantum Metric Tensor and Related 
Topological Phase Transition with a Superconducting Qubit'}, arXiv:1908.06462v1.

\bibitem{tan-erratum} X. Tan, \textit{et al.}, \textit{Erratum: Experimental Measurement of the Quantum Metric Tensor and Related Topological Phase Transition with a Superconducting Qubit}, Phys. Rev. Lett. \textbf{123}, 159902(E) (2019).

\bibitem{gianfrate}  Gianfrate, A., \textit{et al.}, \textit{Measurement of the quantum geometric tensor and of the anomalous Hall drift}, Nature \textbf{578}, 381-385 (2020).

\bibitem{asteria} L. Asteria, D. T. Tran, T. Ozawa, M. Tarnowski, B. S.
Rem, N. Fl\"{a}schner, K. Sengstock, N. Goldman, and C. Weitenberg, \textit{Measuring quantized circular dichroism
in ultracold topological matter}, Nature physics \textbf{15}, 449 (2019).

\bibitem{yu} M. Yu, P. Yang, M. Gong, Q. Cao, Q. Lu, H. Liu, S. Zhang, M. B. Plenio, F. Jelezko, T. Ozawa, et al., \textit{Experimental measurement of the quantum geometric tensor using coupled qubits in diamond}, National Science
Review \textbf{7}, 254 (2020).

\bibitem{kless} R. L. Klees, G. Rastelli, J. C. Cuevas, and W. Belzig, \textit{Microwave spectroscopy reveals the quantum geometric
tensor of topological Josephson matter}, Phys. Rev. Lett. \textbf{124}, 197002 (2020).

\bibitem{liu} Yu,M., Liu, Y., Yang, P. \textit{et al.}  \textit{Quantum Fisher information measurement and verification of the quantum Cram\'er-Rao bound in a solid-state qubit}, npj Quantum Inf \textbf{8}, 56 (2022).
 
\bibitem{yu-quan-ma} Y.-Q. Ma, S.-J. Gu, S. Chen, H. Fan, and W.-M. Liu, \textit{The Euler number of Bloch states manifold and the quantum phases in gapped fermionic systems}, Europhys. Lett. \textbf{103}, 10008 (2013).

\bibitem{kolodrubetz} M. Kolodrubetz, V. Gritsev, and A. Polkovnikov, \textit{Classifying and measuring geometry of a quantum ground state manifold}, Phys. Rev. B \textbf{88} 064304 (2013).

\bibitem{roy} R. Roy, \textit{Band geometry of fractional topological insulators}, Phys. Rev. B  \textbf{90}, 165139 (2014).

\bibitem{lee} C. Lee, et al., \textit{Band structure engineering of ideal fractional Chern insulators}, Phys. Rev. B \textbf{96}, 165150 (2017).

\bibitem{bj-yang} S. Kwon and Bohm-Jung Yang, \textit{Quantum geometric bound and ideal condition for Euler band topology}, arXiv 2311.11577 (2023). 

\bibitem{yang-2015} Lu Yang, Yu-Quan Ma, and Xiang-Gui Li, \textit{Geometric tensor and the topological characterization of the Bloch band in a two-band lattice model}, Physica B \textbf{456} (2015) 359-364.

\bibitem{yu-quan-ma-2}Y.-Q. Ma, \textit{Euler characteristic number of the energy band and the reason for its non-integer values}, arXiv:2001.05946 (2020). 

\bibitem{mera}B. Mera and T. Ozawa, \textit{K\"{a}hler geometry and Chern insulators: Relations between topology and the quantum metric}, Phys. Rev. B. \textbf{104}, 045104 (2021).

\bibitem{ozawa}T. Ozawa and B. Mera, \textit{Relations between topology and the quantum metric for Chern insulators}, Phys. Rev. B. \textbf{104}, 045103 (2021).

\bibitem{milnor} \textit{Characteristic Classes}, Milnor and Stasheff, Princeton University Press and University of Tokyo Press (1974).

\bibitem{lee-book} John M. Lee, \textit{Introduction to Smooth Manifolds}, Second Edition. Springer Science+Business Media New York 2003, (2013).


\bibitem{orientation} Furthermore, note that $\Omega_{12}$ can also be written in a compact form $2\Omega_{12}=d_{\mathcal{M}}={x}\cdot\big(x_{1}\times x_{2}\big)$, with $x_{i}=\partial x/\partial k^i$, representing the projection of the normal vector to the surface $\mathcal{M}_{s,\alpha}\subset S^2$ to the position vector ${x}$ when both are non colinear. From Eq. (\ref{det-N}), $d_{\mathcal{N}}$ has the same sign of $\Omega_{12}$. 

\bibitem{qi} X.L. Qi, Y.S. Wu, S.C. Zhang, \textit{Topological quantization of the spin Hall effect in two-dimensional paramagnetic semiconductors}, Phys. Rev. B \textbf{74} (2006) 045125.


\bibitem{haldane} F. D. M. Haldane, \textit{Model for a quantum Hall effect without Landau levels: Condensed-matter realization of the "parity anomaly"}; Phys. Rev. Lett. \textbf{61} 2015 (1988).

\bibitem{fruchart} \textit{Parallel Transport and Band Theory in Crystals}, Michel Fruchart \textit{et al} 2014 EPL \textbf{106} 60002.

\bibitem{carpentier} Carpentier, D. (2017). \textit{Topology of Bands in Solids: From Insulators to Dirac Matter.} In: Duplantier, B; Rivasseau, V; Fuchs, JN (eds) Dirac Matter. Progress in Mathematical Physics, Vol \textbf{71}. Birkh\"{a}user, Cham. 

\bibitem{madsen-book} \textit{From Calculus to Cohomology}, Madsen, I. and Tornehave, J.; Cambridge University Press (1997).

\end{thebibliography}
\end{document}